\documentclass[PAPER, atlasdraft=false, UKenglish, cernpreprint, texmf, orcidlogo]{atlasdoc}
\usepackage{atlaspackage}
\usepackage{atlasbiblatex}

\usepackage{atlasphysics}
\graphicspath{{logos/}{figures/}}

\AtlasTitle{Study of Particle Fluence Effects on Collected Charge and Depletion Voltage of the ATLAS IBL Planar Pixel Sensors}

\AtlasAbstract{%
After ten years of operation at the LHC, the planar pixel sensors of the innermost barrel layer of the ATLAS Pixel detector have accumulated an average bulk damage fluence in excess of $2\times10^{15}$ 1~MeV-neutrons equivalent/cm$^2$. The macroscopic effects of this radiation are an increase of the sensor leakage current, a loss of charge collection efficiency and an increase of the depletion voltage. Using regular bias voltage scans performed at the beginning and end of each data taking campaign the evolution of the pixel cluster charge and bulk depletion is studied as a function of particle fluence. Results are interpreted with the modelling provided by standalone TCAD and ATLAS Monte Carlo simulation including radiation damage effects. The dependence of the collected charge and the depletion voltage with integrated luminosity are studied through the full period of operation.
}

\AtlasRefCode{PIXE-2025-01}

\PreprintIdNumber{CERN-EP-2026-129}

\AtlasJournal{Journal of Instrumentation}
\AtlasCoverEgroupAnalysisTeam{atlas-PIXE-2025-01-analysis-team@cern.ch}

\hypersetup{pdftitle={ATLAS document},pdfauthor={The ATLAS Collaboration}}

\begin{document}

\maketitle

\tableofcontents

\section{Introduction}
\label{sec:intro}

%
%

Hybrid pixel sensors are the main technology of choice for the innermost tracking layers in the multi-purpose experiments at the LHC collider. They provide precise space points along the trajectories of charged particles in the vicinity of the beam collision region for particle tracking, vertexing and hadronic jet flavour tagging.
The evolution of the charge collected as a function of the applied reverse-bias voltage and the particle fluence tests the processes of charge collection in the Si bulk. These impact the spatial resolution of the reconstructed pixel hits and the identification of pixel clusters with contributions from multiple charged particles in dense hadronic jets, crucial to the construction of physics objects depending on pixel response. In addition, the monitoring of the depletion voltage is a key parameter for a performant pixel operation. As an effect of irradiation, the value of the depletion voltage increases with the integrated luminosity and the particle fluence accumulated on the detector. This increase needs to be regularly monitored and the applied bias voltage adapted for an efficient detector operation.

In ten years of operation, the innermost barrel layer of the ATLAS Pixel detector, the Insertable B Layer (IBL), has accumulated a fluence, $\Phi$, of more than $2\times10^{15}$ 1~MeV-neutrons equivalent (\SI{}{n_{eq}}) per \SI{}{cm^2}. The dependence of the collected charge with the bias voltage is regularly measured during $pp$ collisions in dedicated bias voltage scans. These provide a significant corpus of data to study the evolution of the charge collection process and the depletion voltage over two orders of magnitude of particle fluence values. The observables measured in data are not limited to the depletion voltage but include also the charge collection efficiency and the fraction of collected charge as a function of the depth of charge generation in the Si bulk. Performing this study on the IBL after ten years of operations at the LHC provides the broadest available range of particle fluences  and, therefore, bulk damage effects on the ATLAS pixel sensors.

Results on effects of radiation damage to charge collection and depletion voltage in Si sensors installed around the interaction region at hadron colliders have been published for microstrip~\cite{CDF:2013mfe}, hybrid planar (see chapter~5 of Ref.~\cite{CYP} and Ref.~\cite{CMSDP2025006}) and 3D~\cite{ATLAS:2024ytx} pixel sensors.

In this study, detailed simulations of the electric field profile and of the charge generation and trapping processes provide first principles insights towards understanding the trends in data observables as a function of bias voltage and particle fluence.
These trends are studied in this paper in terms of the electric field profile, the charge carrier drift velocity, their collection time and the trapping effects from radiation induced defects in the Si bulk.

This paper is organised as follows. The IBL planar pixel sensors, their operation and bias voltage scan procedures are briefly presented in Section~\ref{sec:pixels}. Section~\ref{sec:raddamage} discusses the radiation levels for the IBL sensors and the modelling of the effects of radiation damage on the sensors.  Section~\ref{sec:data} presents the data analysis procedures employed in this study and discusses the evolution of the collected charge with fluence, depth of charge generation and bias voltage in comparison to the simulation predictions. The determination of the depletion voltages as function of the integrated luminosity and fluence is the subject of Section~\ref{sec:depl}. Their interpretation through detailed simulations is presented in Section~\ref{sec:tcad}. Section~\ref{sec:conclusions} has the conclusions.


\section{The ATLAS Pixel Detector}
\label{sec:pixels}

%
%

The ATLAS Pixel Detector consists of four concentric cylindrical layers in the barrel region and three layers on either side in the forward region~\cite{Aad_2008}. It operates at the centre of the Inner Detector (ID), which also includes stereo pairs of silicon microstrip detector (SCT) layers and a transition radiation tracker (TRT), inside a 2~T solenoidal magnetic field, in the ATLAS detector~\cite{atlas-detector}. The Insertable B Layer (IBL)~\cite{ibl,ATLAS:2010ojh} is the innermost layer of the ATLAS Pixel Detector and was installed before the start of LHC \RunTwo. It consists of 14 staves installed to form a cylinder around the LHC beam pipe at an average radius of 33.5~\si{\mm}. Each stave supports twenty pixel sensors together with electrical services and cooling. Two distinct sensor technologies are used: slim-edge planar sensors and 3D sensors.

Results on charge collection properties and depletion voltages for IBL 3D sensors as a function of the particle fluence have been recently published~\cite{ATLAS:2024ytx}. In this paper we focus on the IBL planar sensors. These are installed in twelve modules equipping the central section of the IBL staves, covering the longitudinal region, along the beam axis, of $|z|<245$~\si{mm}, in the pseudorapidity range of $|\eta|<1.90$.\footnote{ ATLAS uses a right-handed coordinate system with its origin at the nominal interaction point (IP) in the centre of the detector and the \(z\)-axis along the beam pipe. The \(x\)-axis points from the IP to the centre of the LHC ring, and the \(y\)-axis points upwards. Polar coordinates \((r,\phi)\) are used in the transverse plane, \(\phi\) being the azimuthal angle around the \(z\)-axis. The pseudorapidity, $\eta$, is defined in terms of the polar angle \(\theta\) as \(\eta = -\ln \tan(\theta/2)\).} Staves are inclined such that the average incidence angle of a straight track at $\eta$ = 0 is 14\si{\degr} for the IBL. Contiguous modules overlap to ensure full coverage in the azimuthal angle and for alignment purposes.

\subsection{IBL Planar Sensor Design and Operation}
\label{sec:iblsensors}

IBL planar sensors are $\sim$18~\si{\mm} wide and $\sim$41~\si{\mm} long, in the direction longitudinal to the stave. Sensors have n$^+$-in-n pixels with a pixel pitch of 50~$\mu$m in the transverse and 250~$\mu$m in the longitudinal coordinate.
The n-side segmentation matches in size the FE-I4 front-end (FE) read-out electronics~\cite{fei4} connected via bump-bonds. A guard-ring structure is placed on the p-side. The IBL planar sensors are produced using n-type wafers of 100~mm diameter and 200~$\mu$m thickness, with resistivity in the range of 2--5~k$\Omega$cm~\cite{ibl}.

Two 80$\times$336 pixels FE chips are bump-bonded to a planar sensor. The two central columns of the double-chip sensors are extended to 450~$\mu$m to cover the gap between the two adjacent FE chips. The FE-I4 cell features an analog block performing signal amplification and discrimination and a digital block computing the Time over Threshold (ToT), expressed in units of the 25~ns LHC clock, which is correlated to the detected charge and coded on 4 bits. If the charge collected on a pixel exceeds the IBL dynamic range, set at approximately 30~k$e^-$, an overflow bit is set.  The ToT is tuned to a preset value, corresponding to a given injected charge on reference capacitors in the FE chips. Threshold and noise are determined by analysing the response curve obtained for injected charge values around the set threshold. During most of its operation, the IBL analog threshold was set to 1500~$e^-$. Given the possible drift of the ToT-to-charge response due to operation and irradiation effects, tunings and calibrations are regularly performed during data-taking periods. The ToT response of each FE chip corresponding to various injected charge values are recorded in look-up tables and loaded to the calibration database. The ToT-to-charge conversion for the IBL is performed using these look-up tables in the offline event reconstruction.

Sensors on each stave are organised in module groups of two double-chip planar sensors and share a common bias voltage channel and leakage current measurement. The maximum bias voltage that can be provided by the power supplies to planar sensors is 1000~V, at a maximum current of 8~mA.

The temperature is monitored by negative temperature coefficient (NTC) thermistors mounted on the module flex hybrid that routes signal and power lines between the stave flex hybrid board and the FE-I4 chips. The granularity of the temperature monitoring is the same as with the bias voltage channels. The cooling is provided by a ${\text{CO}}_2$ two-phase system, with the coolant being circulated in titanium pipes embedded in the stave structure~\cite{ibl}. The cooling was set to $-$\SI{10}{\degreeCelsius} at the start of operation in 2015, and increased to \SI{+15}{\degreeCelsius} and \SI{+5}{\degreeCelsius} in 2016 to reduce the effect of the total ionising dose (TID) on the FE-I4 chips, which increases the leakage current of the transistors and therefore the power consumption of the chip~\cite{CYP}. After that period, cold operation was resumed with a cooling set point of $-$\SI{20}{\degreeCelsius} for most of the operation during \RunTwo and \RunThr, corresponding to temperatures of $\simeq$\,${-}$\SI{13}{\degreeCelsius} measured on the module flex hybrids during $pp$ collision data taking. The detector was kept cold during the LHC end-of-year technical stops and the long shutdown in between \RunTwo and \RunThr, except for short periods due to maintenance and interventions.

The operation bias voltage, $V_{\mathrm{bias}}$, applied to the IBL planar sensors was increased progressively with time: from 80~V in 2015 to 400~V in 2018, during \RunTwo, and from 450 in 2022 to 650~V in 2026, during \RunThr, to ensure sensor operation well above the depletion voltage. Regular I--V scans were performed during periods without beam to monitor the level of the leakage current and the evolution of the breakdown voltages with the accumulated radiation. Moreover, dedicated bias voltage scans were run a few times during each data taking period. %

\subsection{Bias Voltage Scans}

Collision events were collected at closely spaced values of bias voltages during dedicated, short bias voltage scans performed in \RunTwo at the beginning of the 2016 run, with only 4.5~fb$^{-1}$ of integrated delivered luminosity, and at the end of the 2016, 2017 and 2018 LHC $pp$ collision periods, corresponding to 45, 93 and 155~fb$^{-1}$ of integrated luminosity. In \RunThr  bias voltage scans were performed at the beginning and end of 2022, 2023 and 2024, corresponding to 194, 228 and 345~fb$^{-1}$ of integrated luminosity, and at the beginning, mid and end of 2025, corresponding to 345, 410 and 470~fb$^{-1}$ of integrated luminosity. In order to study the charge collection in irradiated sensors, the evolution of the charge collected in clusters for hits associated to reconstructed particle tracks is studied as a function of the applied bias voltage. This gives information about the depletion in the sensor Si bulk and the charge collection properties. Data from the voltage scans are compared with the predictions of the Radiation Damage simulation, discussed in Section~\ref{sec:raddamage}, with electric field maps corresponding to the fluence value estimated for the corresponding data sets.

The pixel bias voltage ranges used in the voltage scans have needed to evolve over the years due to the increasing depletion voltage arising from the bulk damage to the sensors with particle fluence.
At the low end of the voltage range, the decrease of charge collection efficiency makes the measured cluster charge progressively affected by analog threshold effects. This is due to cases in which part of the charge collected on neighbouring pixels for particles traversing the active Si volume between pixel implants is lost because the collected charge on one of the pixels is below threshold. The most probable value of the pixel cluster charge in these cases is no longer representative of the collected charge. The lower bias voltage points, progressively more affected by these charge losses, are gradually excluded in the study. At the high end of the voltage scan range, bias voltages are applied only after verifying that the currents driven by the sensors are low enough for safe operation. Guided by these principles, the voltage range adopted in the bias voltage scans has evolved from 7 -- 80~V at the beginning of 2016 to 170 -- 650~V in 2025.


\section{Fluence and Radiation Damage Simulation}
\label{sec:raddamage}

%
%

\subsection{Radiation levels for the IBL sensors}

At the radius of the IBL the primary source of radiation damage is due to primary collision products. The values of particle fluence on the detector surface are estimated from leakage current data (see chapter~3 of Ref.~\cite{CYP} and Ref.~\cite{ATLAS:2021gld}). The conversion factor between delivered integrated luminosity and the fluence on planar sensors is obtained by averaging the values obtained as a function of the $z$ position weighted by the $z$ distribution of the measured particle track impacts on the IBL surface of planar sensors in the data samples used for the analysis. This yields a value of $(5.40 \pm 0.35) \times 10^{12}$~\SI{}{n_{eq}}~cm$^{-2}$/fb$^{-1}$, where the quoted uncertainty accounts for systematic uncertainties from the leakage current analysis discussed in Ref.~\cite{ATLAS:2021gld}. The leakage current analysis has shown the fluence longitudinal profile to be less uniform than those predicted by the FLUKA~\cite{BATTISTONI201510} and \GEANT~\cite{GEANT4:2002zbu} simulations. At the radius of the IBL modules the estimated fluence decreases by $\sim$30\% moving from the detector centre to the outer edges of the planar sensor modules~\cite{ATLAS:2021gld}. There is no evidence of fluence asymmmetry in the azimuthal angle $\phi$.

The IBL planar sensors received an estimated average fluence of $8.6 \times 10^{14}$~\SI{}{n_{eq}}~cm$^{-2}$ during \RunTwo and $1.67 \times 10^{15}$~\SI{}{n_{eq}}~cm$^{-2}$ in the four years of \RunThr LHC operations up to the end of 2025. The total particle fluence is expected to approach $3 \times 10^{15}$~\SI{}{n_{eq}}~cm$^{-2}$ by the end of \RunThr in 2026, remaining within the sensor specifications set at the time of the IBL design~\cite{ATLAS:2010ojh}.

\subsection{Radiation damage simulation}
\label{sec:raddamsim}

The ATLAS Collaboration has developed a detailed simulation of radiation damage effects in the pixel sensors~\cite{ATLAS:2019epb}. The pixel radiation damage simulation is currently used in the production of ATLAS Monte Carlo (MC) samples for physics analyses and it is used to predict the expected charge collection efficiency as a function of integrated luminosity, to optimise operation conditions, and to calibrate the reconstruction tools.

A radiation damage digitiser models the process of charge collection and signal generation on the readout nodes. It computes the signals induced on the read-out electrodes from a detailed simulation of the drift of charge carriers produced by ionising particles, modelled using the Bichsel model~\cite{RevModPhys.60.663}, taking into account radiation effects in the sensor. The carriers' speed is calculated using the product of their mobility~\cite{JACOBONI197777} and the estimated electric field in the Si bulk. This is simulated using Silvaco\footnote{www.silvaco.com/tcad} TCAD\footnote{Technology Computer Aided Design} tools. Effective defects are used in the simulation to reproduce the effects of radiation damage. The electric field profile in planar sensors after irradiation is computed as discussed below in Section~\ref{sec:EFieldRad}. This profile is used to calculate the average charge deflection due to the Lorentz force in the ATLAS solenoidal magnetic field. The combined effect of the electric and magnetic fields is used to estimate the time for a carrier to reach the respective collecting electrode. If the expected collection time is larger than an exponentially distributed trapping time with average value equal to $\tau = 1/(\beta \Phi)$, where $\beta$ is the trapping constant and $\Phi$ the fluence, then the carrier is considered trapped and its induced signal is calculated using a precomputed weighting potential~\cite{Ramo} map of the sensor.

\subsection{Electric Field in irradiated sensors}
\label{sec:EFieldRad}

\begin{figure}[ht!]
\begin{center}
\includegraphics[width=.495\textwidth]{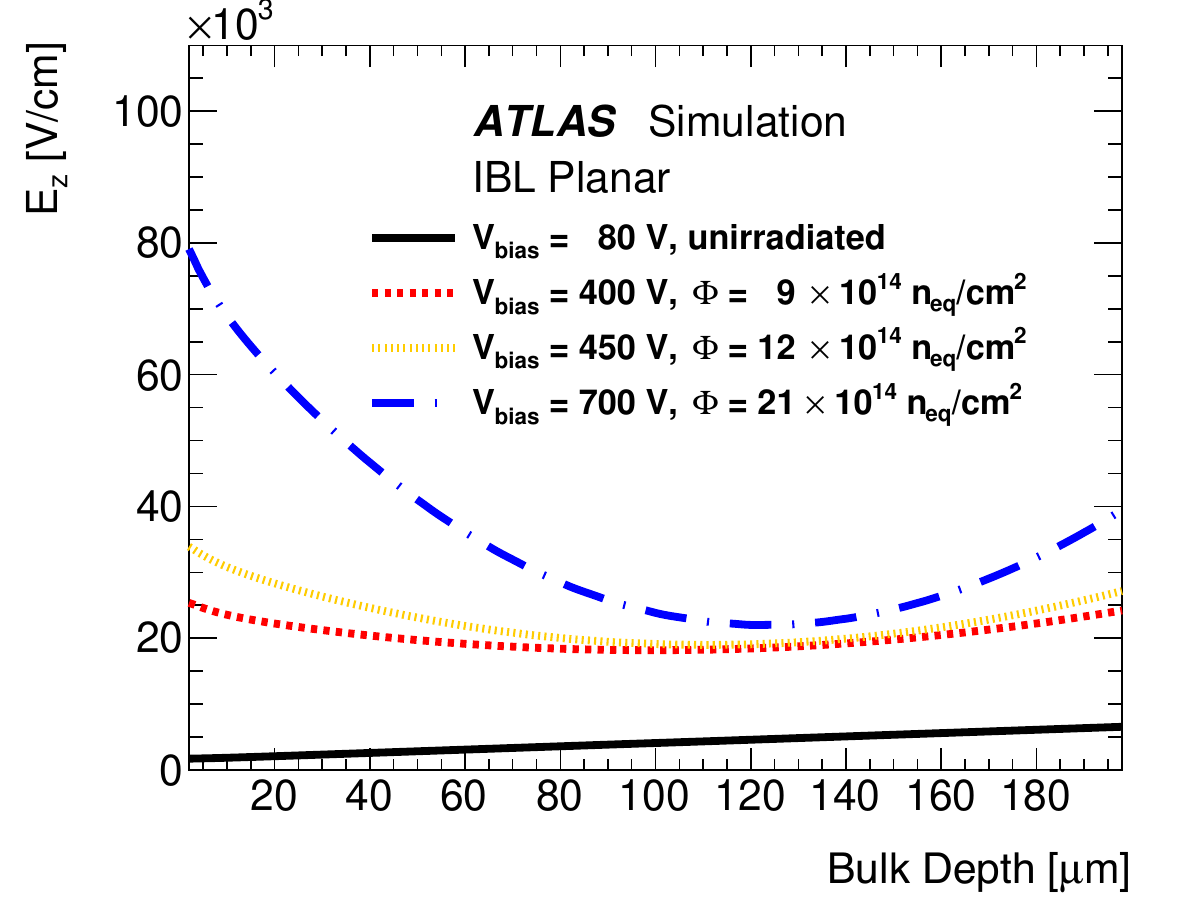}
\end{center}
\caption{TCAD simulated electric field profile of IBL planar sensors. The electric field projection along the bulk is shown as a function of the bulk depth. The electrons drift towards smaller depth values. The different fluences correspond to the values reached at the beginning of Run~2 (unirradiated), at the end of Run~2 ($9\times10^{14}$~\SI{}{n_{eq}}~cm$^{-2}$, for 161~fb$^{-1}$ integrated luminosity), at the end of 2023  ($12\times10^{14}$~\SI{}{n_{eq}}~cm$^{-2}$, for 230 fb$^{-1}$ integrated luminosity), and the projection for the end of Run~3 in 2026 ($21\times10^{14}$~\SI{}{n_{eq}}~cm$^{-2}$). The bias voltages correspond to the data-taking conditions. \label{fig:efieldprofile}}
\end{figure}

The observed collected charge as a function of bias voltage ($V_{\mathrm{bias}}$) in the voltage scans is the result of the interplay of charge carrier drift and charge trapping. The charge carrier velocities are the product of the local electric field $\vec{E}$ and the mobility $\mu$ (which depends on the electric field itself). The electric field depends on the space charge distribution  at the operating depletion voltage. At the level of irradiation experienced by the innermost layers of the ATLAS Pixel detector, ${\mathcal{O}}(10^{15})$~\SI{}{n_{eq}}~cm$^{-2}$, the distribution of space charge strongly depends on the occupation of radiation-induced deep defects that gives rise to an electric field profile characterised by a deep minimum close to the mid-plane, and large maxima at the junction and ohmic side. Radiation damage effects are simulated in TCAD using the Chiochia model~\cite{Chiochia:2004qh,CHIOCHIA2006,Adam2016}, based on the original work of Ref.~\cite{bib:DP}. The evolution of the density and cross sections of the defects includes a sub-linear dependence of damage with accumulated fluence~\cite{Cartiglia:2019eut}. The electric field profiles obtained from these simulations provide the input for the ATLAS radiation damage digitizer~\cite{ATLAS:2019epb}. Typical results for the electric field profiles for an IBL planar pixel sensor at different values of irradiation fluence and operating voltage are shown in Figure~\ref{fig:efieldprofile}.

A significant change in the electric field profile in the sensor bulk at the start of operations and after an irradiation equivalent to that received by the IBL sensors at the end of 2023 is visible. At the minimum of the electric field, the carrier velocities can be below their saturation value ($\sim$~\SI{100}{\micro\meter}/ns) if the applied bias voltage is not high enough. This increases the probability for charge carriers to get trapped by the deep defects induced by bulk damage.

\subsection{Systematic Uncertainties on Charge Collection Predictions}
\label{sec:Systs}

The systematic uncertainties in the charge collection predictions of irradiated IBL pixel sensors are due to four main sources. These are: i) the luminosity-to-fluence conversion, ii) the electric field profile, iii) the charge trapping constant, $1/\tau$, and iv) the sensor operating conditions (temperature, etc). Of these,  the trapping constant gives the largest contribution. At a fluence of $10^{15}$~\SI{}{n_{eq}}~cm$^{-2}$, a variation of the proportionality factor, $\beta$, between the trapping constant $1/\tau$ and the fluence $\Phi$ by $\pm$ 30\%, justified by the dispersion of data available in literature~\cite{ATLAS:2019epb}, induces a relative change of the most probable value (MPV) of the collected cluster charge by $\pm$12\%. The uncertainty on the luminosity-to-fluence conversion of  $(5.40 \pm 0.35) \times 10^{12}$~\SI{}{n_{eq}}~cm$^{-2}$/fb$^{-1}$ contributes a relative change of $\pm$3\%. The description of the sensor operating conditions in the MC is responsible for an additional relative uncertainty estimated at $\pm$2\%. Finally, the electric field uncertainties, computed by modifying the parameters of the effective states of the TCAD radiation damage model (energy, density and carrier cross section, see Ref.~\cite{ATLAS:2019epb}), yield a relative change of the cluster charge MPV of only $\pm$1\%.  The luminosity-to-fluence conversion and trapping constant uncertainties, being proportional to the fluence, increase with the amount of radiation damage. The error bands associated with the simulation predictions in this study reflect the impact of these uncertainties on the IBL sensor observables.


\section{Charge Collection Measurements}
\label{sec:data}

%
%

Regular voltage scans during $pp$ collisions facilitated the study of collected charge as a function of both the applied voltage and of the bulk depth at which the charge was created, and at various radiation levels.
The results are systematically compared with the prediction of the radiation damage simulation. Results on charge collection are presented as a function of fluence and depth of charge generation in this section and as a function of bias voltage in Section~\ref{sec:depl}. Charge collection vs.\ bias voltage are then interpreted using simulation results in Section~\ref{sec:tcad}.

\subsection{Data Samples and Analysis Procedure}
\label{sec:perf-datasets}

This study is based on $pp$ collision data collected during \RunTwo and \RunThr operations from 2015 to 2018 at $\sqrt{s} = 13$~TeV and from 2022 to the end of 2025 at $\sqrt{s} = 13.6$~TeV, respectively.

The software infrastructure developed for the reconstruction process of real detector data and MC simulation are presented in Ref.~\cite{ATLAS:2024vdo}. In the reconstruction, pixel clusters are built from adjacent pixels reporting charge above a preset analogue threshold, as discussed in Section~\ref{sec:iblsensors}. In this study, only pixel clusters reconstructed in the IBL planar sensor modules operating at the nominal bias voltage and associated with a reconstructed charged-particle track (\enquote{hits-on-track}) are considered.

Particle tracks from collision events are selected by requiring transverse momentum, \pT, in excess of 0.7~GeV for the determination of charge and in excess of 3~GeV for the studies of depth of charge generation discussed in Section~\ref{sec:depth}.  In addition, track selection requires at least two hits in the Pixel layers, with at least one in the IBL, a minimum of seven hits summed across the Pixel and SCT layers and track extrapolation compatible with the position of the reconstructed primary vertex~\cite{ATL-PHYS-PUB-2019-015}.
Pixel clusters generating hits associated with these selected tracks are considered in this study. In addition, pixel clusters shared or split between two or more particles tracks, as well as those containing one or more pixels with the charge overflow bit set, are discarded. Since the cluster charge depends on the particle track length in the active Si bulk, the charge is rescaled by $\cos \alpha$, where $\alpha$ is the particle's angle of incidence on the sensor surface.

The studies of charge collection efficiency in Sections~\ref{sec:cce} and \ref{sec:depth} use collision events fulfilling single hadronic jet triggers~\cite{TRIG-2016-01, TRIG-2022-01}. The bias voltage scans, due to the limited amount of events collected at each bias voltage setting, use all events in the main physics stream~\cite{TRIG-2022-01}. Samples of minimum bias and dijet QCD events were generated at 13 and 13.6~TeV with \Pythia[8]~\cite{Bierlich:2022pfr} and reconstructed after applying the ATLAS detector simulation. This simulation includes the pixel radiation damage digitiser, discussed in Section~\ref{sec:raddamsim}, for the signal formation. Reconstructed charged-particle tracks in simulation are selected using the same criteria as used for data. In addition, simulated particle tracks are reweighted to reproduce the distribution of the incidence angle $\alpha$ on the IBL sensor surface measured in data.
An extensive validation of the radiation damage digitiser was performed, where the predictions for IBL planar sensors are compared with collision data. Results for charge collection are presented in the following sections.

\subsection{Charge Collection Efficiency vs.\ Fluence}
\label{sec:cce}

\begin{figure}[ht]
\centering
\subfloat[]{
\includegraphics[width=0.45\textwidth]{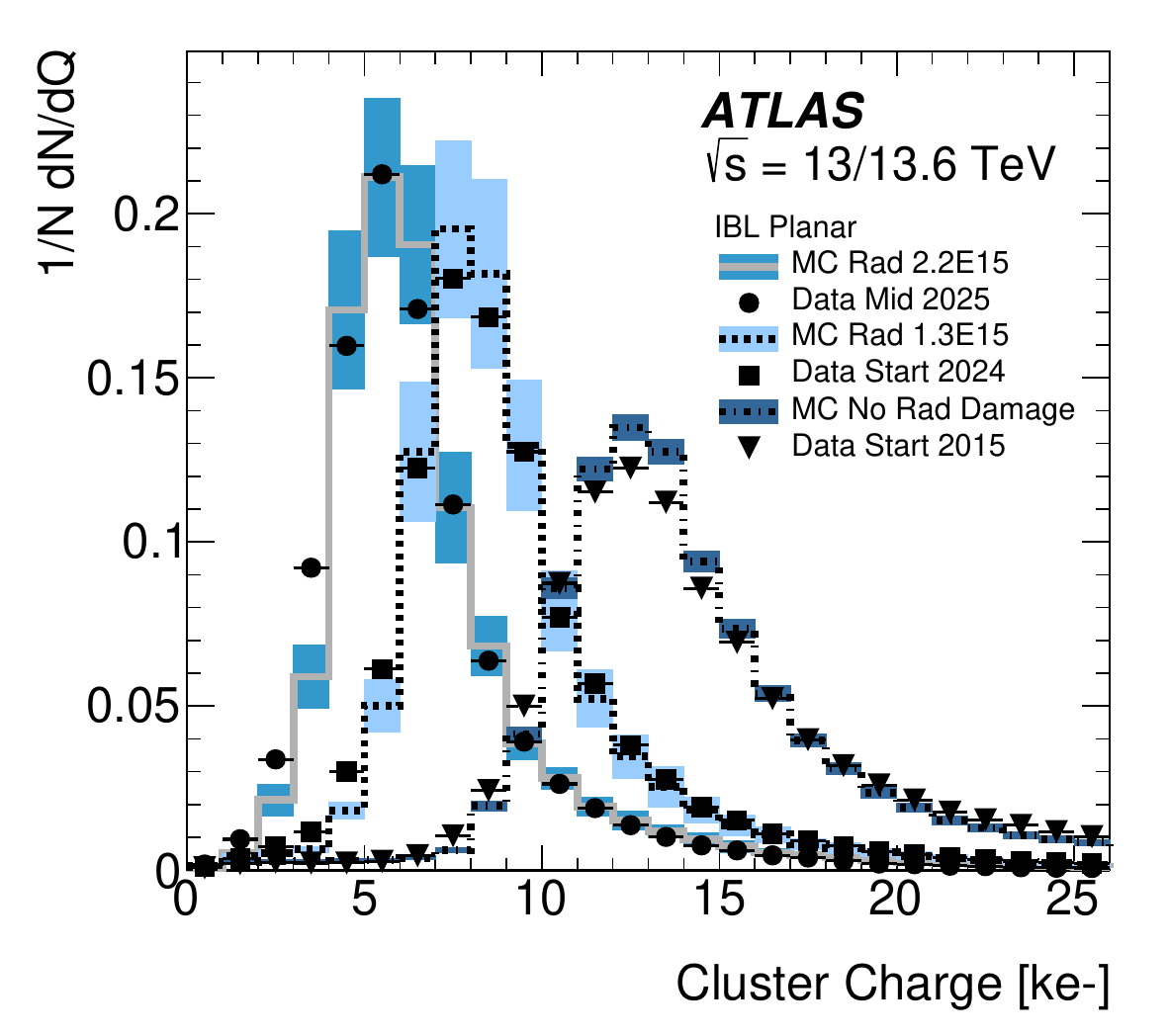}
\label{fig:Planar_Cluster_Charge}
}
\subfloat[]{
\includegraphics[width=0.475\textwidth]{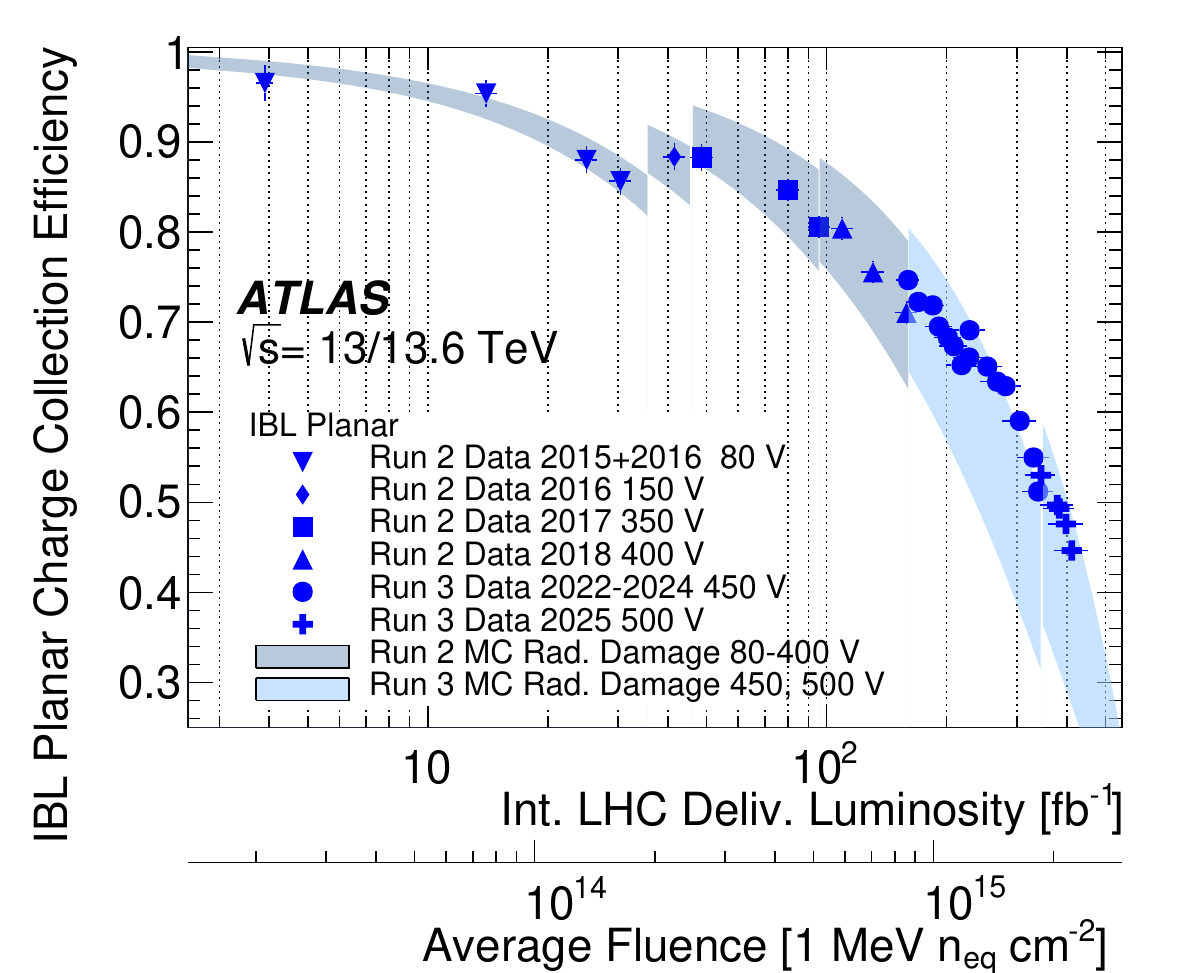}
\label{fig:Planar_CCE}
}
\caption{Charge collection in IBL and radiation damage: (Left) Distribution of cluster charges in IBL planar sensors from data taken at the start of the 2015 (filled triangles) and 2024 (filled squares) and mid-2025 (filled circles) compared with the predictions from radiation damage simulation corresponding to the estimated  average fluence.  The shaded regions give the range of the uncertainties in simulation. (Right) Charge collection efficiency as a function of the integrated delivered luminosity and average fluence for IBL planar sensors for data and the ATLAS radiation damage simulation from the beginning of Run 2. The points represent the data and the bands the simulation predictions with their uncertainties. Predictions for the evolution until the end of \RunThr are also given. \label{fig:Planar_Charge}}
\end{figure}

The measured cluster charge in IBL planar sensors and the evolution of the charge collection efficiency (CCE), defined as the ratio of the MPV of the collected cluster charge normalised to the value measured at the beginning of sensor operation in 2015, are shown in comparison with the simulation predictions in Figure~\ref{fig:Planar_Charge}. Only data taken in the first run following a ToT calibration are considered for determining the CCE as a function of integrated luminosity and accumulated fluence.

The sudden increases in the charge collection efficiency at the beginning of each year are due to changes in the operational parameters, in particular increases in the bias voltage applied to the IBL sensors. The simulation predictions agree with the measurements using collision data over almost two orders of magnitude of radiation fluence. For fluence values above $\simeq 1.5 \times 10^{15}$~\SI{}{n_{eq}/cm^2}, data may start exhibiting a moderately larger charge collection efficiency compared to simulation predictions. As already mentioned, a sub-linear dependence of charge trapping with accumulated fluence may be expected in heavily irradiated sensors~\cite{Cartiglia:2019eut}. This effect could be tested on data collected at higher fluences at the end of Run~3.

\subsection{Charge Collection Efficiency vs.\ Charge Generation Depth}
\label{sec:depth}

In planar sensors, charge trapping due to Si bulk radiation damage reduces the charge collection efficiency as the depth of charge generation increases. The evolution of the charge collection efficiency as a function of the charge generation depth is key to the understanding of the observed charge vs.\ bias voltage in the voltage scans. This can be studied by comparing data and simulation at different fluences and bias voltage values.
\begin{figure}[ht!]
\begin{center}
\begin{tabular}{cc}
\hspace*{-0.5cm}
\includegraphics[width=0.495\textwidth]{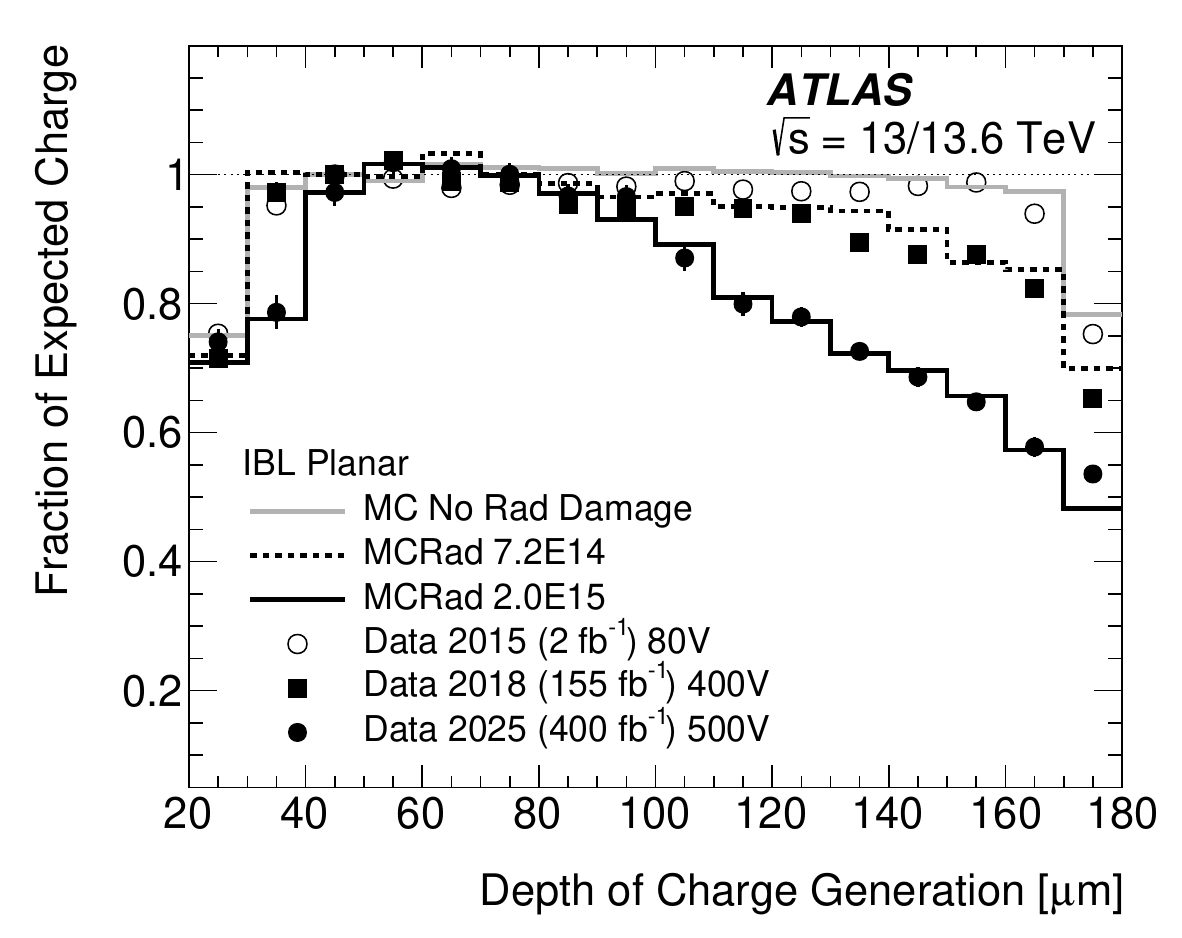} &
\includegraphics[width=0.495\textwidth]{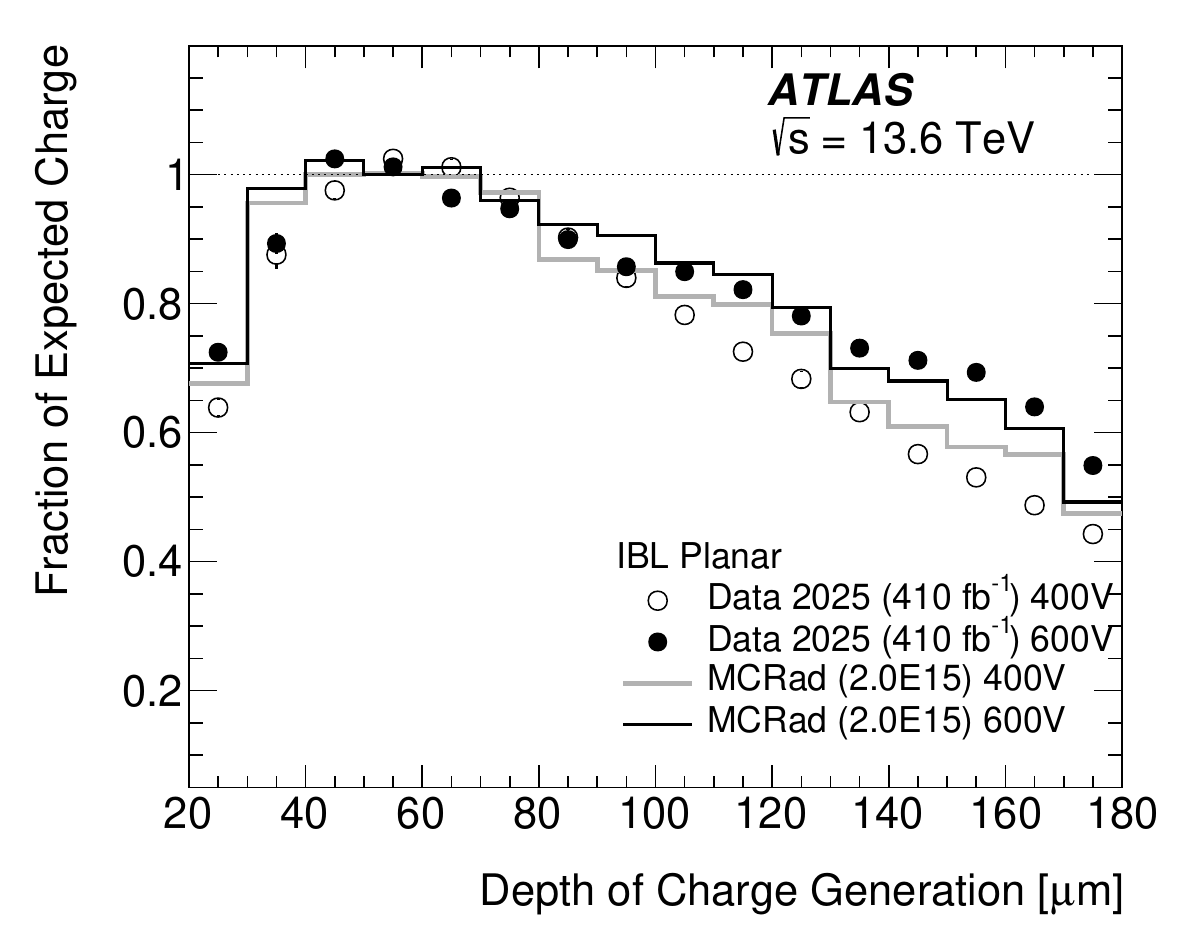}\\
\end{tabular}
\caption{Charge collection efficiency as a function of the estimated depth of charge generation for IBL planar sensors in $pp$ collision data. The left panel shows the evolution from the start of \RunTwo in 2015 (2~fb$^{-1}$ of integrated luminosity shown by the open points), at the end of \RunTwo in 2018 (155~fb$^{-1}$ of integrated luminosity shown by the filled squares) and in \RunThr in mid 2025 (400~fb$^{-1}$ of integrated luminosity shown by the filled points). The right panel shows the evolution with the bias voltage (at 400~V and at 600~V shown by the open and filled points, respectively) in mid 2025 (410~fb$^{-1}$ of integrated luminosity). The predictions from radiation damage simulation for matching average fluence conditions at the position of the IBL modules used in the analysis are shown by the histograms. The efficiency is computed by dividing the fraction of the total cluster charge deposited on the pixels ordered from the extrapolated point of track entrance in the Si substrate to that of track exit in the longitudinal projection by the fraction of the track segment below each pixel to the total track length in the Si bulk. The drops observed at the two ends of the depth range are due to track extrapolation resolution effects. \label{fig:ChargeVsDepth}}
\end{center}
\end{figure}
The measurement method is based on the correlation between the depth of charge generation and the position of each individual pixel in the cluster for tracks traversing the sensitive Si thickness at shallow incidence angles~\cite{LARI2004349,Aad_2008}. By knowing the track angle of incidence and the positions of the track entrance and exit on the Si bulk it is possible to determine the average depth of the track segment below each pixel. Then, the charge collection efficiency as a function of depth is estimated by dividing the fraction of the total cluster charge collected on each individual pixel by the fraction of the track segment below each pixel to the total track length in the Si bulk. For this method to be reliable, two conditions must be fulfilled. First, the distance between the points of track entrance and exit must be large compared to the pixel pitch, so that the charge is deposited uniformly below several adjacent pixels. Second, the extrapolated track position on the detector surface must have a resolution that is small compared to the pixel pitch. Given the ATLAS Pixel detector geometry and pixel pitch, these conditions are met in collision data using the longitudinal position for inclined tracks, depositing charge on three or more pixels along the longitudinal, long pitch, coordinate. In order to optimise the resolution, only tracks with $\pT > 3$~GeV are considered. With these selections, the typical track extrapolation resolution on the surface of the IBL sensors is 150~$\mu$m, to be compared with a longitudinal pixel pitch of 250~$\mu$m. The distributions of the charge collection efficiency as a function of the depth of charge generation obtained at 80, 400 and 500~V after $2\times10^{13}$~\SI{}{n_{eq}/cm^2}, $7.2\times10^{14}$~\SI{}{n_{eq}/cm^2} and $1.9\times10^{15}$~\SI{}{n_{eq}/cm^2}, respectively, and those at 400 and 600~V, after $1.9\times10^{15}$~\SI{}{n_{eq}/cm^2}, are shown in Figure~\ref{fig:ChargeVsDepth} for data and radiation damage simulation. Charge collection inefficiencies from charge trapping are related to the depth of charge generation. These results indicate that the radiation damage simulation gives a good modelling of the charge collection efficiency as a function of the depth, for bias voltage values around the depletion point, at different fluence values.

\subsection{Charge Collection vs.\ Bias Voltage}
\label{sec:chargeHV}

The bias voltage scans provide detailed information about charge collection properties vs.\ bias voltage and particle fluence. In addition, the optimisation of the pixel response in terms of hit-on-track efficiency and spatial resolution depends on the operating bias voltage being above the depletion point. This is regularly monitored and the modelling of its evolution with the integrated luminosity is important for setting the operating conditions before the start of each period of LHC operations.

The bias voltage scans performed in \RunTwo and \RunThr from 2016 to the end of 2025 test the sensor response at estimated average particle fluences ranging over two orders of magnitude, from $2.4\times10^{13}$~\SI{}{n_{eq}/cm^2} to $2.5\times10^{15}$~\SI{}{n_{eq}/cm^2}.

The evolution of the cluster charge vs.\ bias voltage in Figure~\ref{fig:IBLChargeVsHV} probes the charge collection through the sensor bulk and allows us to determine the ``depletion voltage''. For an undepleted reverse-biased junction Si detector, the collected charge is expected to increase $\propto \sqrt{V}$, where $V$ is the applied bias voltage, in proportion to the thickness of the depletion region. An undamaged detector sees the collected charge increase until the Si bulk is fully depleted. This value of the applied bias voltage is commonly referred to as ``depletion voltage''.
At that point the charge saturates and becomes independent of the applied bias voltage. This saturation is observed in the upper left panel of Figure~\ref{fig:IBLChargeVsHV} for data from the voltage scan performed at the start of 2016 after only 4.5~fb$^{-1}$ of integrated luminosity. In a damaged Si detector, the amount of collected charge continues to increase, roughly $\propto V$, above the point of full Si depletion due to the reduction of the charge trapping effect with the increasing charge carrier velocity. In the analysis of the ATLAS pixel data, the depletion voltage is defined as the value of applied voltage at which the transition between the $\propto \sqrt{V}$ and $\propto V$ regimes is observed. This definition has already been adopted in the analysis of the IBL 3D sensors~\cite{ATLAS:2024ytx}.
This simplified picture of the charge collection properties with bias voltage gives a qualitative description of the trends of the collected charge evolution in the voltage scans up to fluences of $\cal{O}\mathrm{(10^{15})}$ \SI{}{n_{eq}/cm^2}. However, the TCAD simulations discussed in Section~\ref{sec:EFieldRad} show that the evolution of the collected charge with the electric field over the relevant range of values of applied bias voltage, depth in the Si bulk and fluence are more complex than in this simplified picture and a detailed analysis is necessary. The results of the analysis of the charge collection vs.\ bias voltage data are presented in the next section and their interpretation based on TCAD simulations in Section~\ref{sec:tcad}.


\section{Charge Collection, Bias Voltage and Depletion}
\label{sec:depl}


%
%

\begin{figure}[ht!]
\begin{center}
\begin{tabular}{ccc}
\includegraphics[width=.325\textwidth]{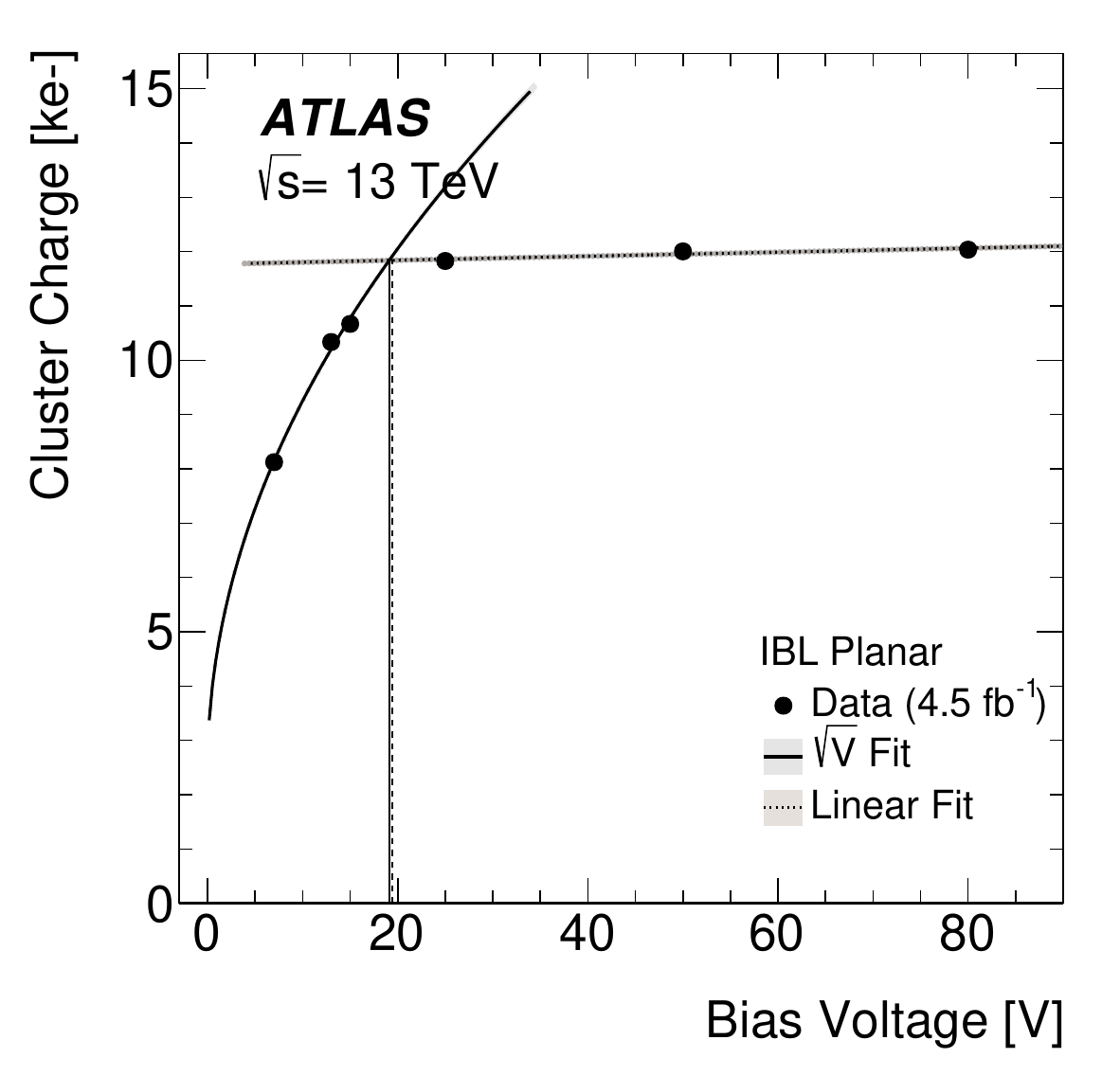} &
\includegraphics[width=.325\textwidth]{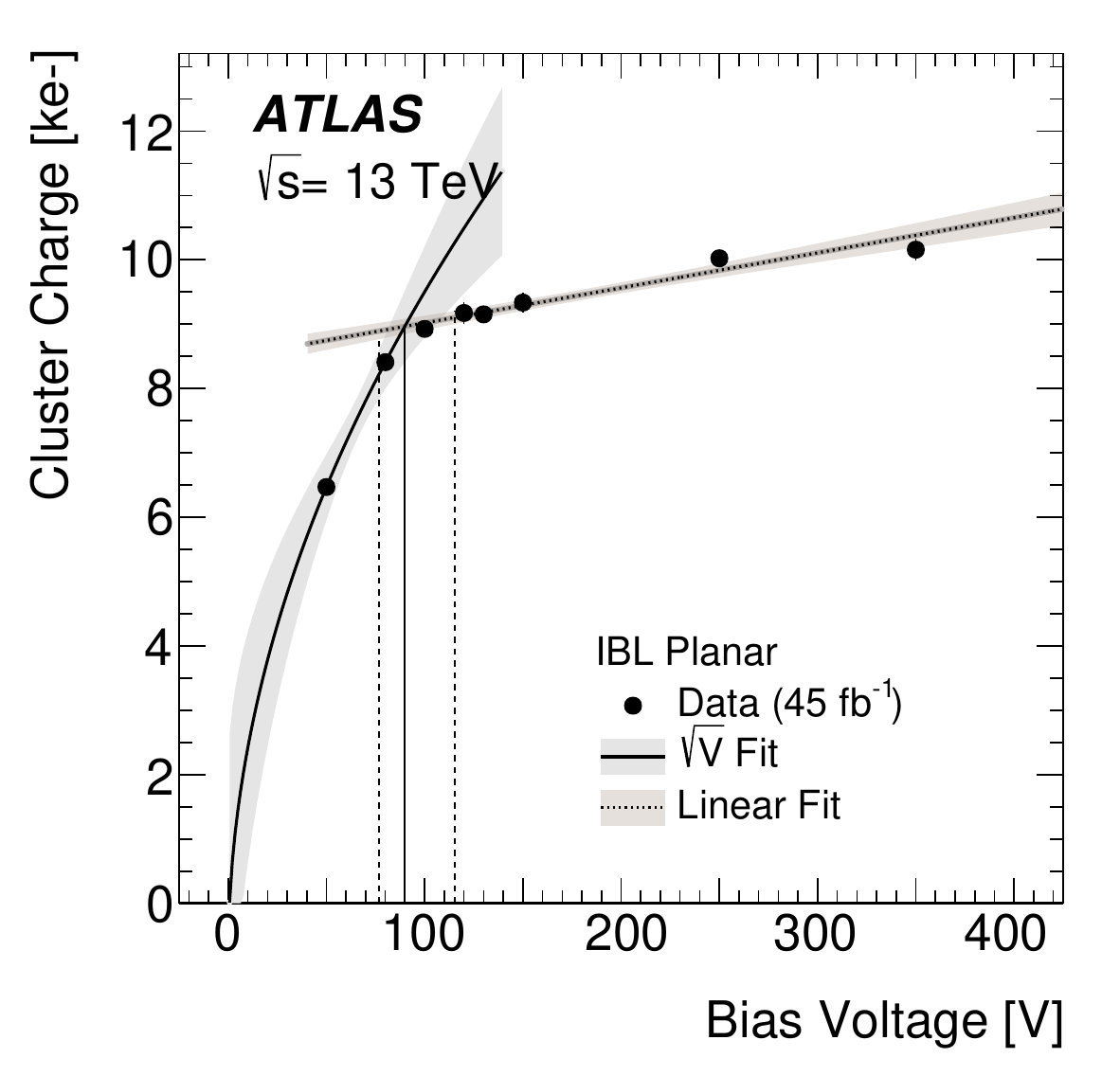} &
\includegraphics[width=.325\textwidth]{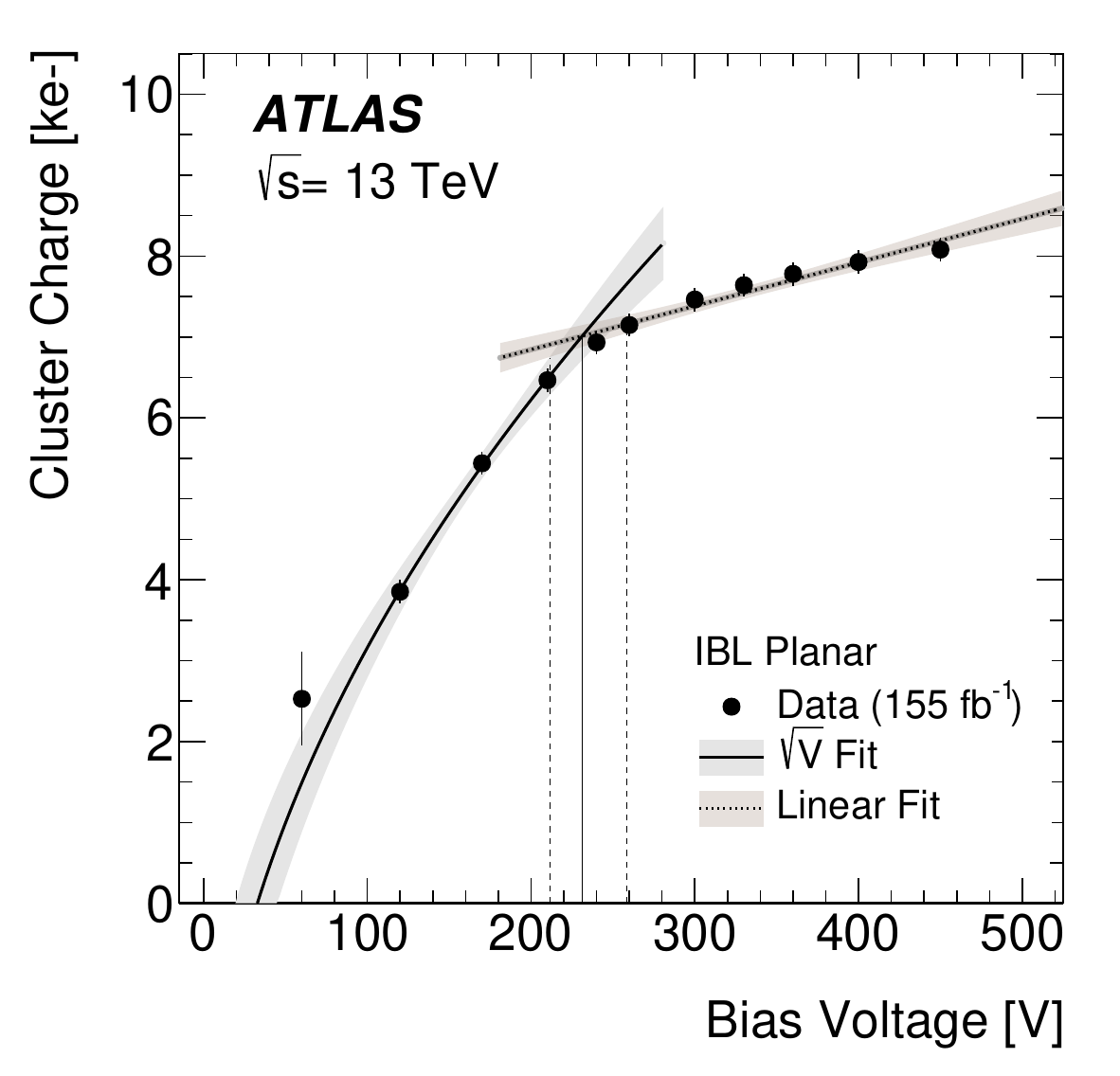} \\
\includegraphics[width=.325\textwidth]{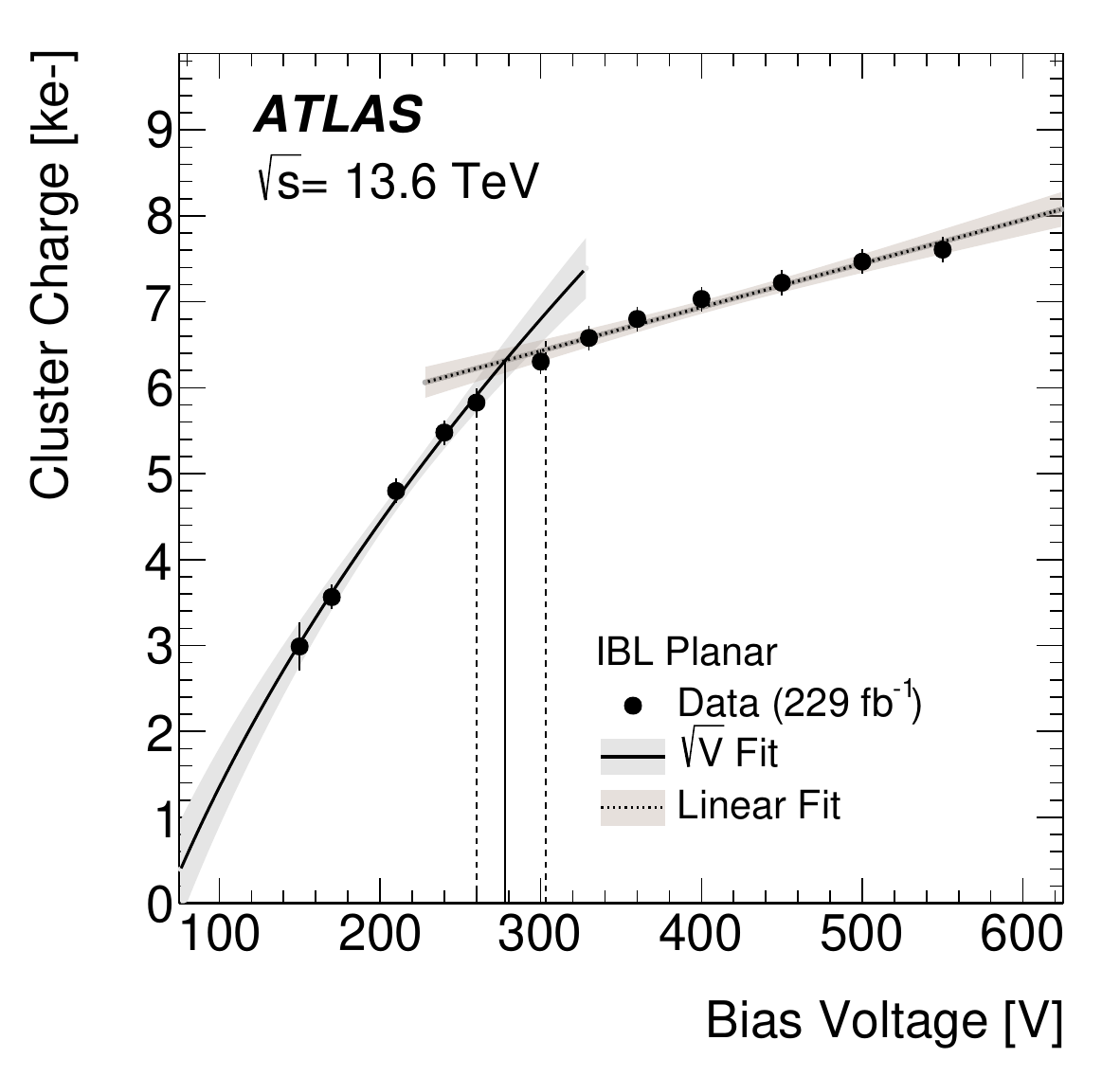} &
\includegraphics[width=.325\textwidth]{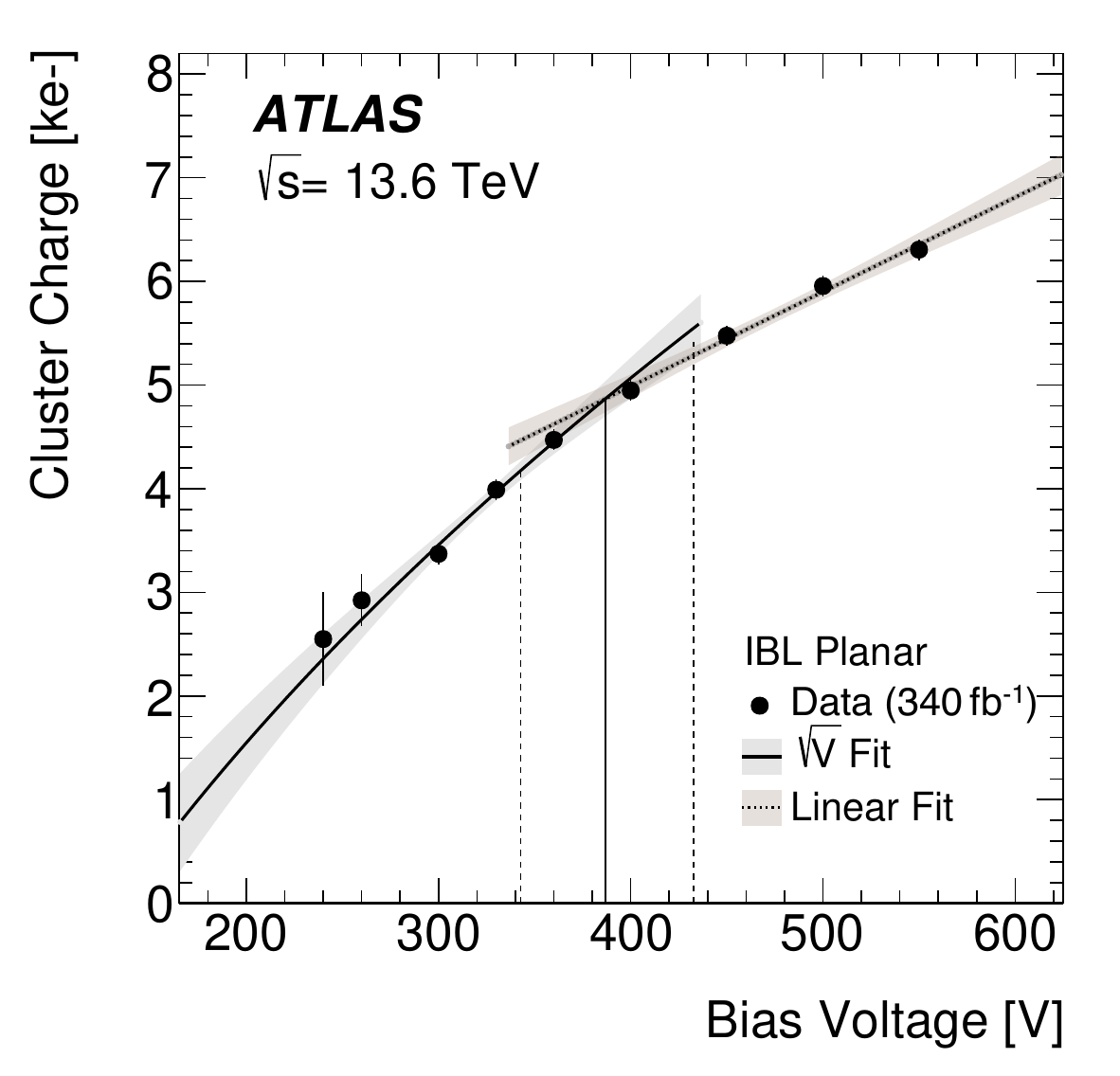} &
\includegraphics[width=.325\textwidth]{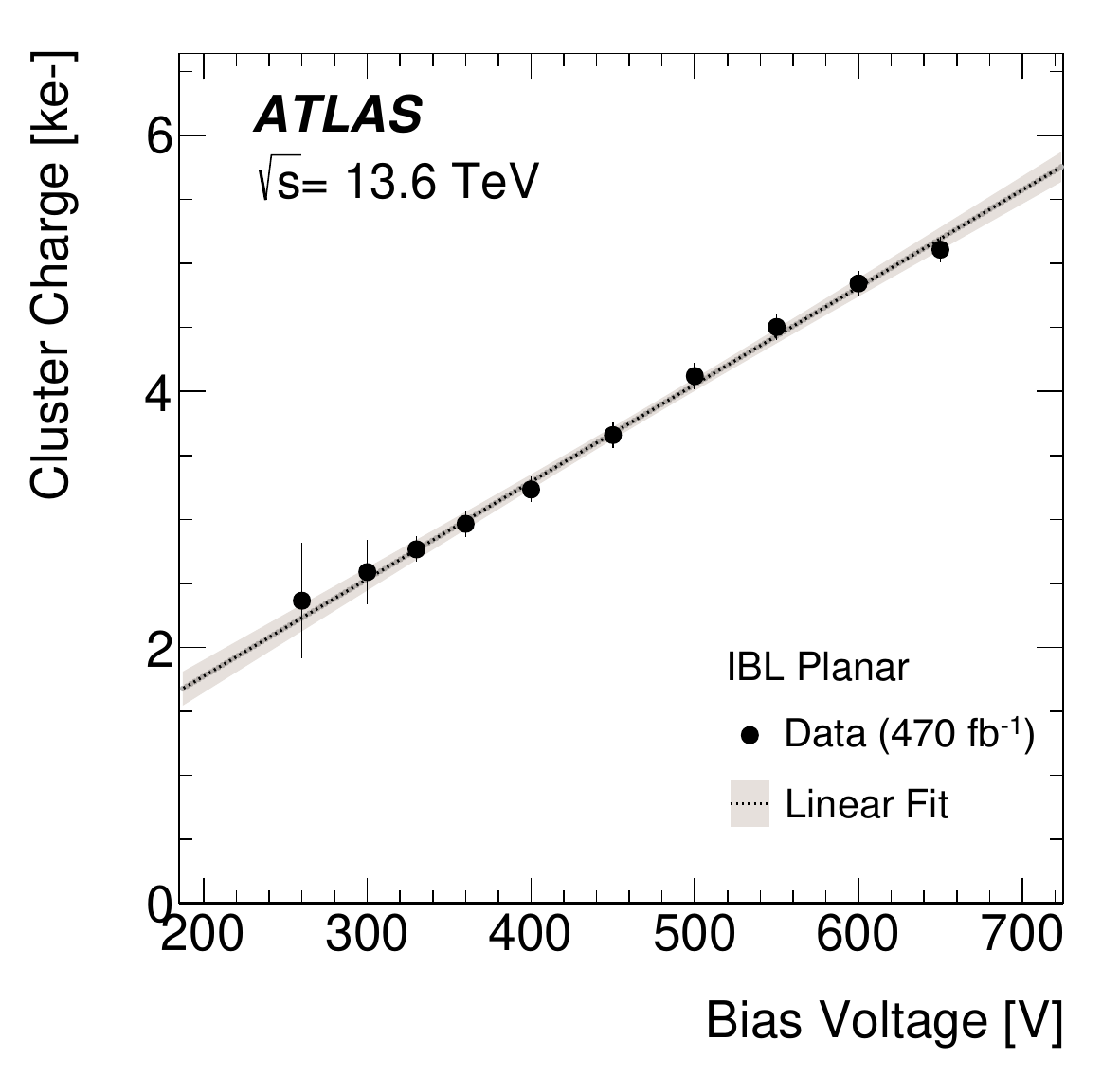} \\
\end{tabular}
\end{center}
\caption{Most probable values of the charge collected in IBL planar sensor clusters associated to reconstructed particle tracks in $pp$ collisions as a function of the bias voltage applied to the sensors, for data recorded in Run~2 (upper plots) and Run~3 (lower plots) from the start of the 2016 run (upper left) to the end of that in 2025 (lower right) at integrated LHC delivered luminosities from 4.5 to 470~fb$^{-1}$. The vertical solid line indicates the value of the fitted depletion voltage with the $\pm~1~\sigma$ interval given by the dashed lines. \label{fig:IBLChargeVsHV}}
\end{figure}

The evolution of the cluster charge with aplied bias voltage is shown in Figure~\ref{fig:IBLChargeVsHV} for data recorded in Run~2 and Run~3 from the start of the 2016 run to the end of that in 2025 at integrated LHC delivered luminosities ranging from 4.5 to 470~fb$^{-1}$, corresponding to fluences from $2.4 \times 10^{13}$~\SI{}{n_{eq}/cm^2} to $2.5 \times 10^{15}$~\SI{}{n_{eq}/cm^2}.

Following the arguments discussed in Section~\ref{sec:chargeHV}, the depletion voltage is defined as the value of applied voltage at which a transition between the $\propto \sqrt{V}$ and $\propto V$ regimes is observed in the evolution of the cluster charge vs.\ the bias voltage. This value is determined by performing an iterative $\chi^2$ fit of a linear $a + b V$ function and a $c + d \sqrt{V}$ function to the voltage scan results, as already done for the analysis of the IBL 3D sensors~\cite{ATLAS:2024ytx}. The points fitted with the linear and the $\sqrt{V}$ functions, from the upper and lower ends of the scan range, are repartitioned until the configuration corresponding to the minimum $\chi^2$ is found. The continuous and dashed lines in Figure~\ref{fig:IBLChargeVsHV} represent the fitted functions modelling the increase in charge collection efficiency in the regimes below and above the depletion point, respectively. The depletion voltage value corresponds to the point of intersection of the two functions indicated by the continuous vertical line. The asymmetric uncertainty is computed by determining the minimum and maximum values of the region associated with the overlap of the error bands for the two fitted functions corresponding to the 68\% confidence level.

With increasing radiation damage, the fitted value of the depletion voltage increases and the slope of the charge evolution above depletion increases while that of region below depletion decreases.
At the fluence of $2.5 \times 10^{15}$~\SI{}{n_{eq}/cm^2}, corresponding to the integrated delivered luminosity of 470~fb$^{-1}$ reached at the end of the 2025 LHC $pp$ run, the slopes of the charge increase with voltage in the two regions merge into a single curve that is approximately linear, as shown in the lower right panel of Figure~\ref{fig:IBLChargeVsHV}.

As discussed in Section~\ref{sec:tcad}, this is interpreted as due to the electrons produced in the deeper part of the bulk having collection times larger than the trapping time for all bias voltage values used in the scans. Under these conditions, the depletion voltage cannot be extracted from the change in the functional shape of the collected charge vs.\ bias voltage. The operational bias voltage is chosen based on the evolution of the hits-on-track efficiency with the voltage.

\begin{figure}[htb!]
\begin{center}
\includegraphics[width=0.525\textwidth]{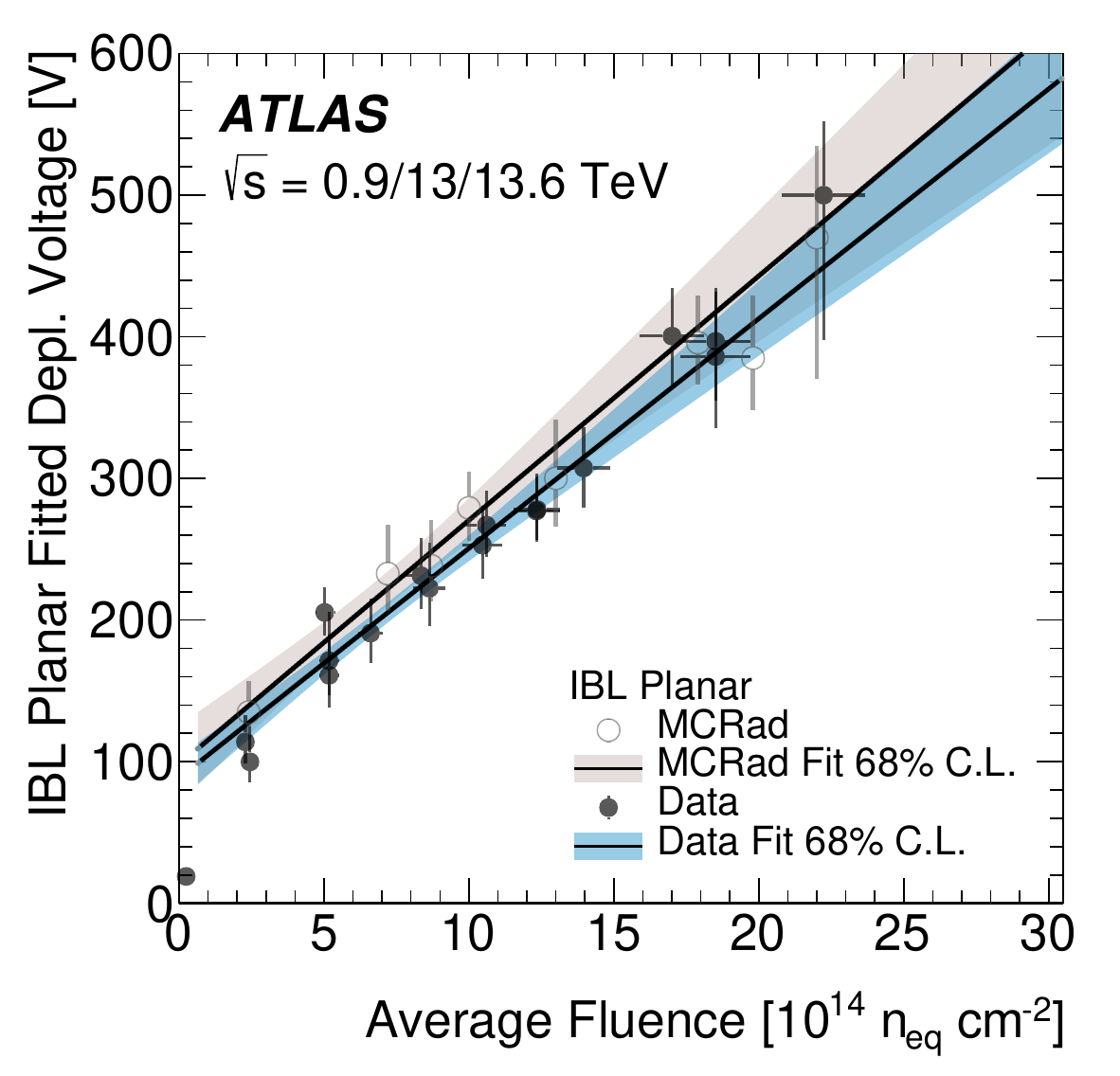}
\caption{Fitted values of the depletion voltage for IBL planar sensors as a function of the average fluence. Values are extracted from fits to the cluster charge as a function of the applied bias voltage in dedicated bias voltage scans conducted typically at the start and end of each year run from 2016 to the mid of the 2025 run and shown by the filled points. The open points show the predictions from radiation damage simulation. The black and the grey line indicates the result of a linear fit over the range of fluence values after the non-linear regime after Si type inversion for data and simulation, respectively. The results shown here indicate that the fitted depletion voltage for IBL planar sensors scales linearly with the received fluence. \label{fig:IBLDeplVsFluence}}
\end{center}
\end{figure}

The depletion voltage values from the voltage scans performed in \RunTwo and \RunThr up to mid-2025 are summarised in Figure~\ref{fig:IBLDeplVsFluence} as a function of the integrated delivered luminosity and average particle fluence with a comparison to the radiation damage simulation predictions.

Since the radiation effects in the run periods are dominant over annealing in the end-of-year LHC stops, the depletion voltage increase can be well modelled linearly with the integrated luminosity and fluence as shown in the fits to both data and simulation in Figure~\ref{fig:IBLDeplVsFluence}. The linear behaviour of depletion vs.\ fluence was already observed in CDF microstrip sensors~\cite{CDF:2013mfe}. These fits provide us with predictions of the depletion voltage by the end of \RunThr, when an estimated 520~fb$^{-1}$ of integrated luminosity will have been delivered by the LHC at the ATLAS interaction region. These depletion voltage predictions are 540$\pm$45~V and 580$\pm$70~V from data and simulation, respectively. IBL planar sensors are biased at 650~V for their last year of operation in 2026.

\begin{figure}[ht!]
\begin{center}
\begin{tabular}{ccc}
\hspace*{-0.5cm}
\includegraphics[width=.325\textwidth]{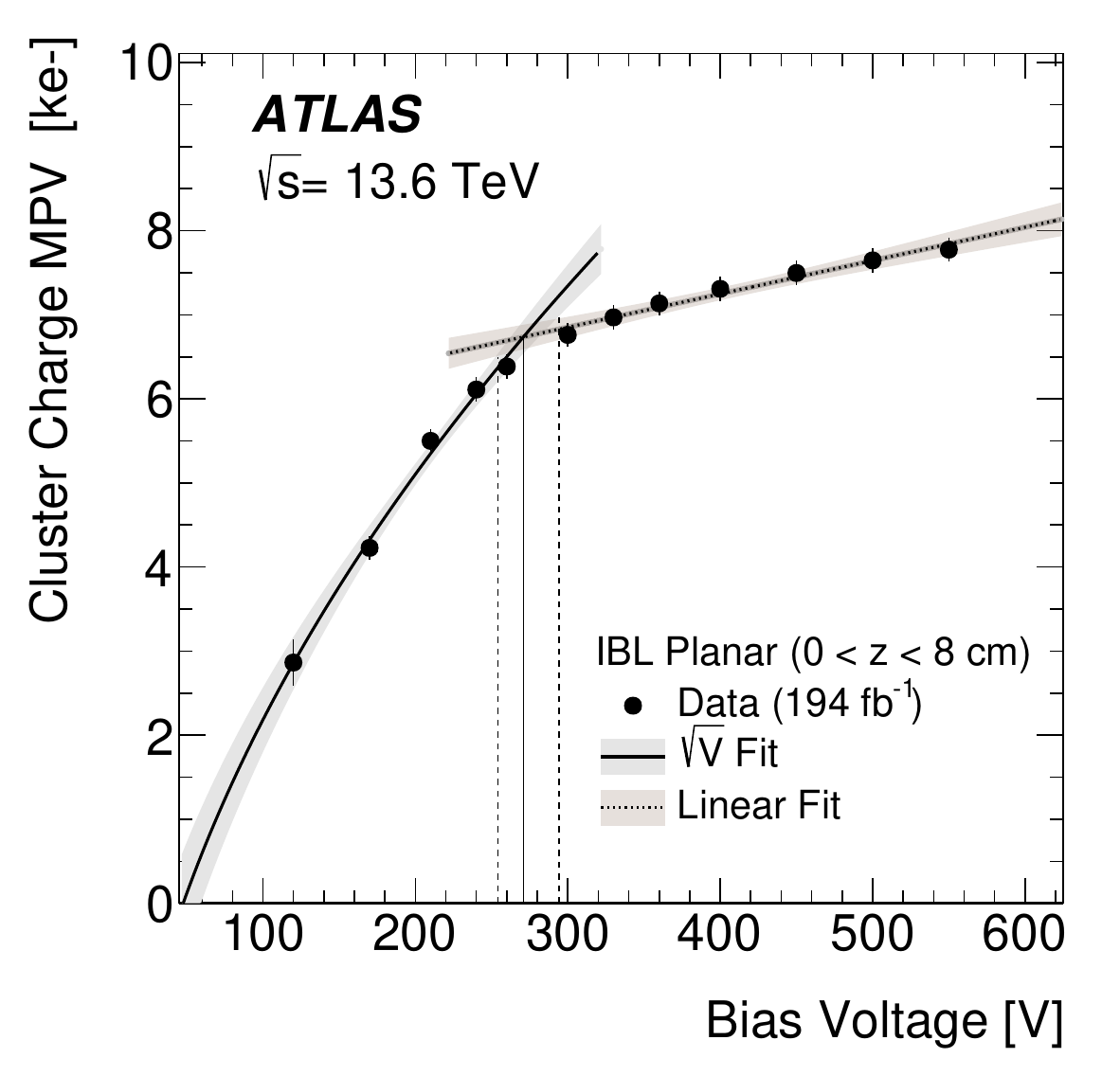} &
\includegraphics[width=.325\textwidth]{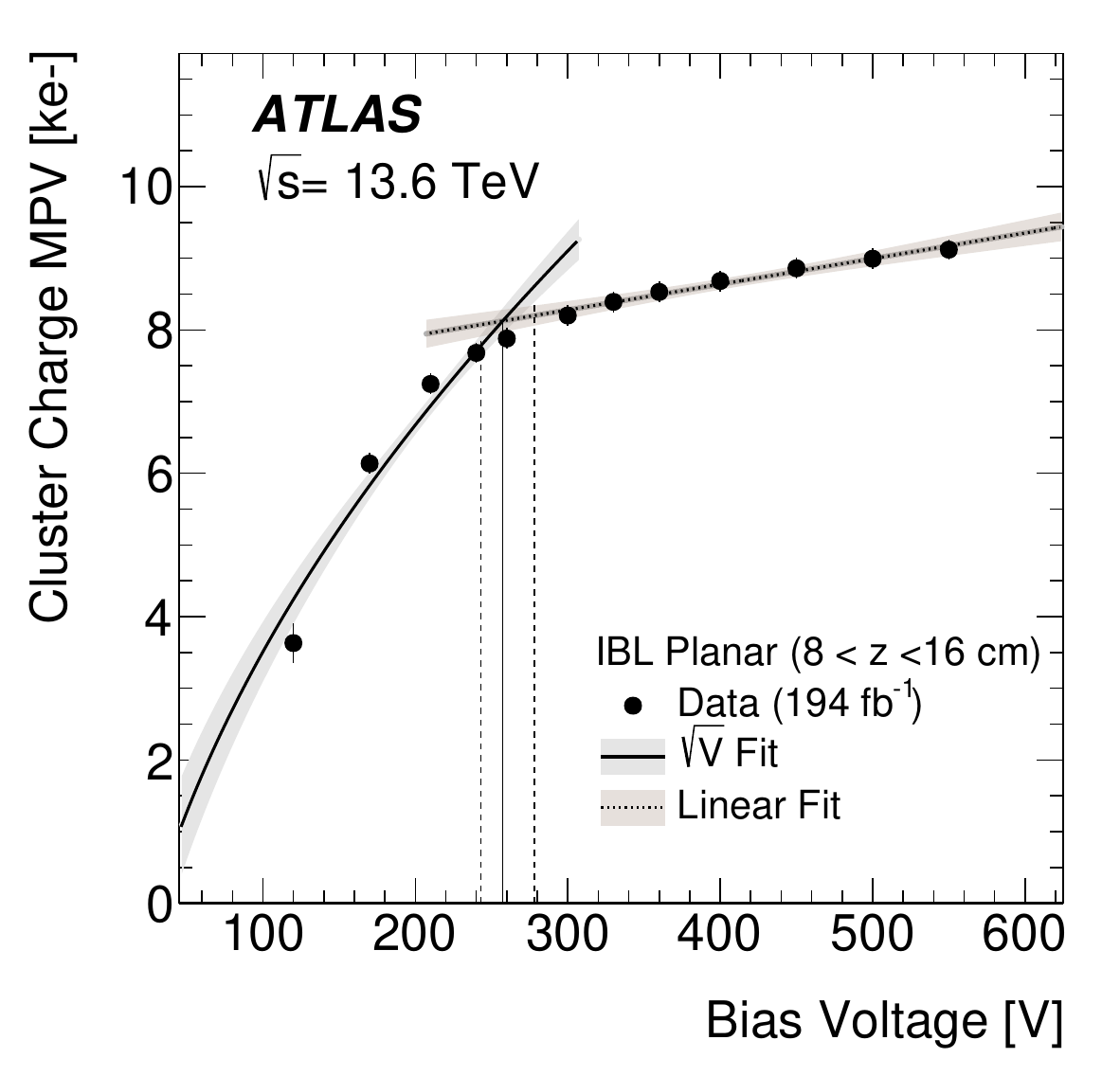} &
\includegraphics[width=.325\textwidth]{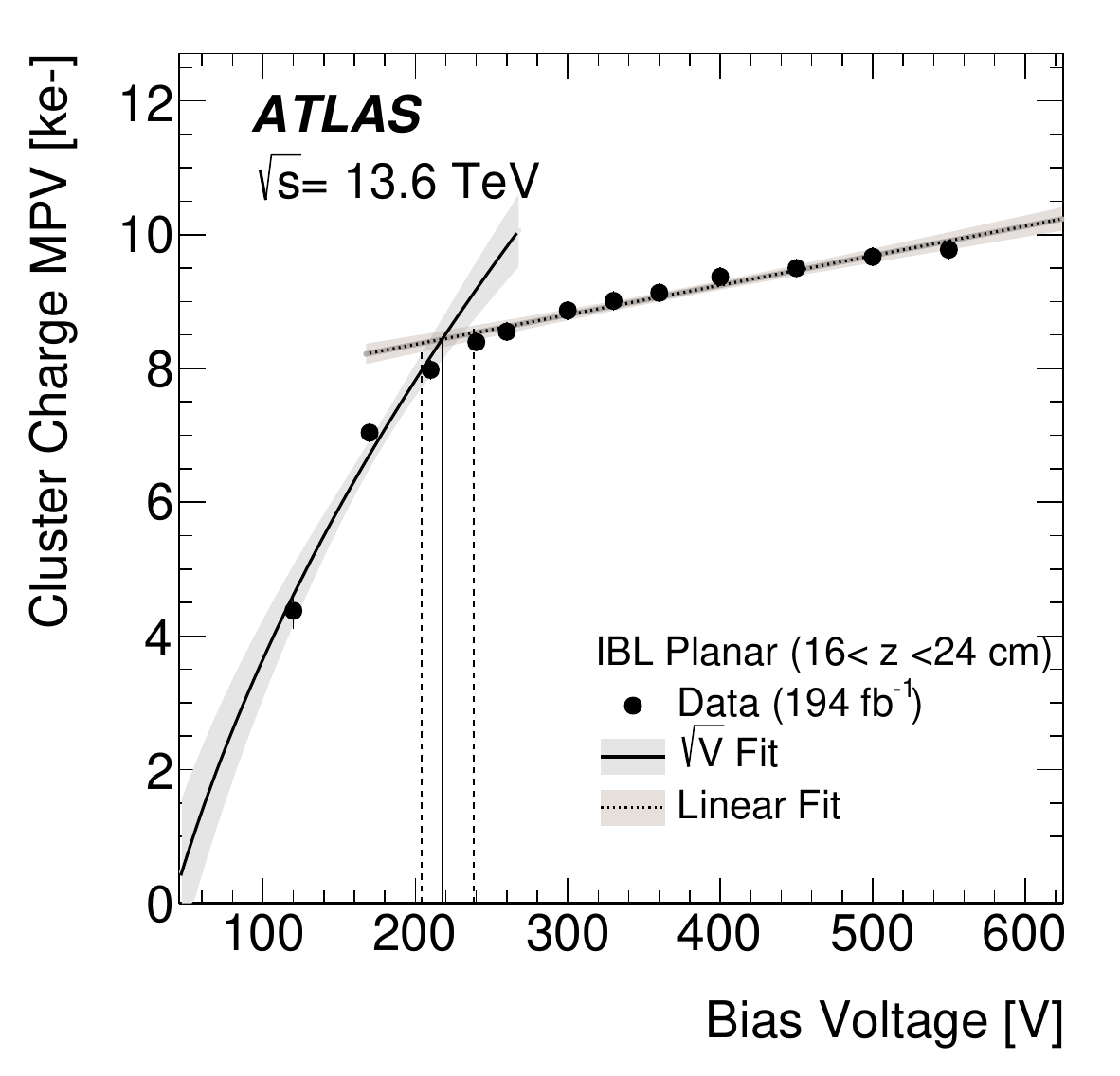} \\
\end{tabular}
\end{center}
\caption{Most probable values of the charge collected in IBL planar sensor clusters associated to reconstructed particle tracks in $pp$ collisions as a function of the bias voltage applied to the sensors, for data recorded at the end of the 2022 run after an integrated luminosity on the IBL of 194 fb$^{-1}$. The three panels show the distributions for clusters reconstructed in three longitudinal module sections from the detector centre to the outer ends (left) for clusters at longitudinal coordinate 0$<|z|<$ 8~cm, (centre) 8$<|z|<$16~cm, and (right) 16$< |z| <$ 24~cm, corresponding to decreasing average fluence values. The depletion voltage values extracted from the fits to the $\sqrt{V}$ and linear regimes, indicated by the solid vertical lines, show their decrease with the lower fluence, when moving from the detector centre to its outer ends. \label{fig:ChargeVsHVZbins}}
\end{figure}

\begin{figure}[htb!]
\begin{center}
\includegraphics[width=0.525\textwidth]{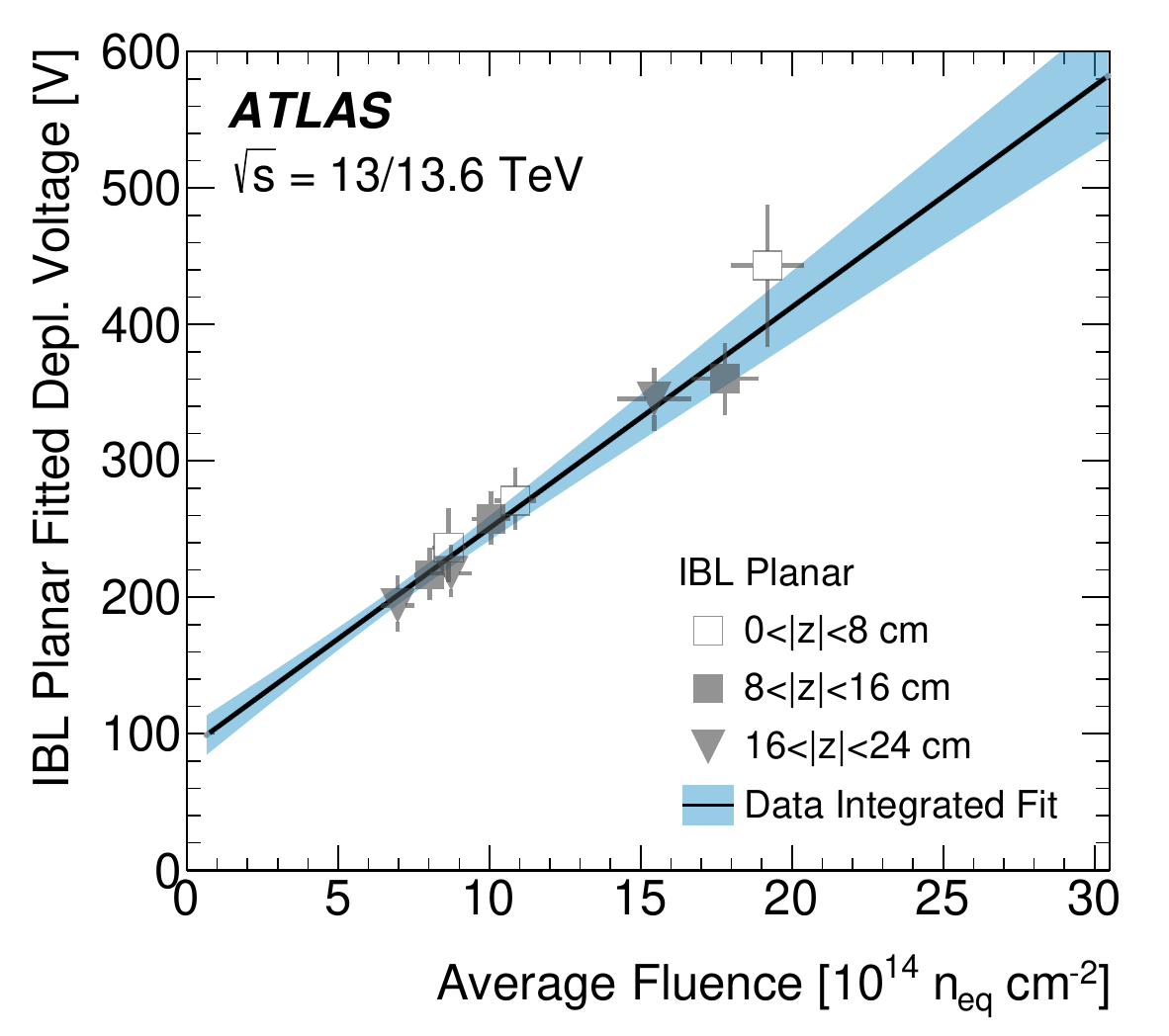} \\
\caption{Fitted values of the depletion voltage for IBL planar sensors as a function of the average fluence. Values are extracted from fits to the cluster charge as a function of the applied bias voltage in dedicated bias voltage scans for data collected at the end of the 2018, 2022 and 2024 runs (with 154, 194 and 343~fb$^{-1}$ of integrated luminosity on the IBL, respectively) over three detector sections symmetric in z, shown by the grey triangles and squares. Given the longitudinal profile of the fluence distribution on the IBL layer, these sections correspond to different average fluence values received by the corresponding IBL sensors. The coloured area indicate the 68\% confidence level envelope of the linear fit to the data, integrated over the full length of the IBL planar modules, from Figure~\ref{fig:IBLDeplVsFluence}. \label{fig:IBLDeplZVsFluence}}
\end{center}
\end{figure}

The particle fluence on the IBL varies along the longitudinal coordinate. It is interesting to study the fitted depletion voltage along the $z$ coordinate and relate it to the average  estimated module fluence. Results are shown in Figure~\ref{fig:ChargeVsHVZbins} with the depletion voltage decreasing while moving from the centre to the edges of the planar sensor section of the detector modules for one scan. Results obtained at the end of \RunTwo in 2018 (161~fb$^{-1}$), and in \RunThr at the end of 2022 (194~fb$^{-1}$) and 2024 (343~fb$^{-1}$) are summarised in Figure~\ref{fig:IBLDeplZVsFluence}. The compatibility of the values extracted from this study on sections of the IBL modules with those obtained using the full IBL modules at different integrated luminosities is tested by comparing these results with the linear fit to the data, integrated over the full length of the IBL planar modules, where good agreement between the two sets of measurements is observed.


\section{Charge Collection and Depletion Voltage in Simulation}
\label{sec:tcad}

%
%

The results obtained from the analysis of the bias voltage scans in data can be interpreted through detailed simulations of electric field and charge collection. Standalone TCAD simulations give first principles insights into the observed dependence of the collected charge on the bias voltage and fluence of an irradiated IBL planar pixel sensor.

\begin{figure}[ht!]
\begin{center}
\begin{tabular}{cc}
\includegraphics[width=.45\textwidth]{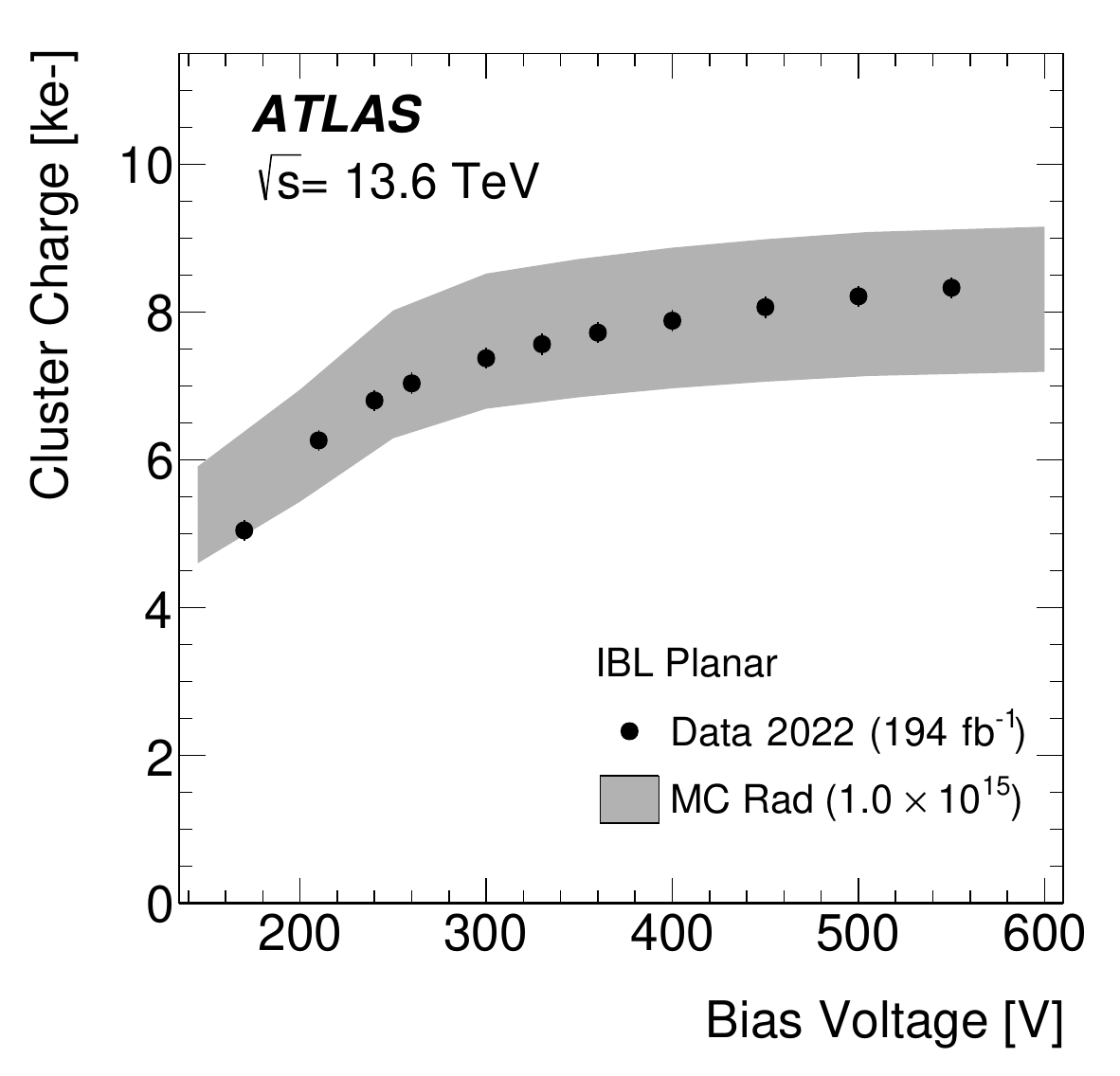} &
\includegraphics[width=.45\textwidth]{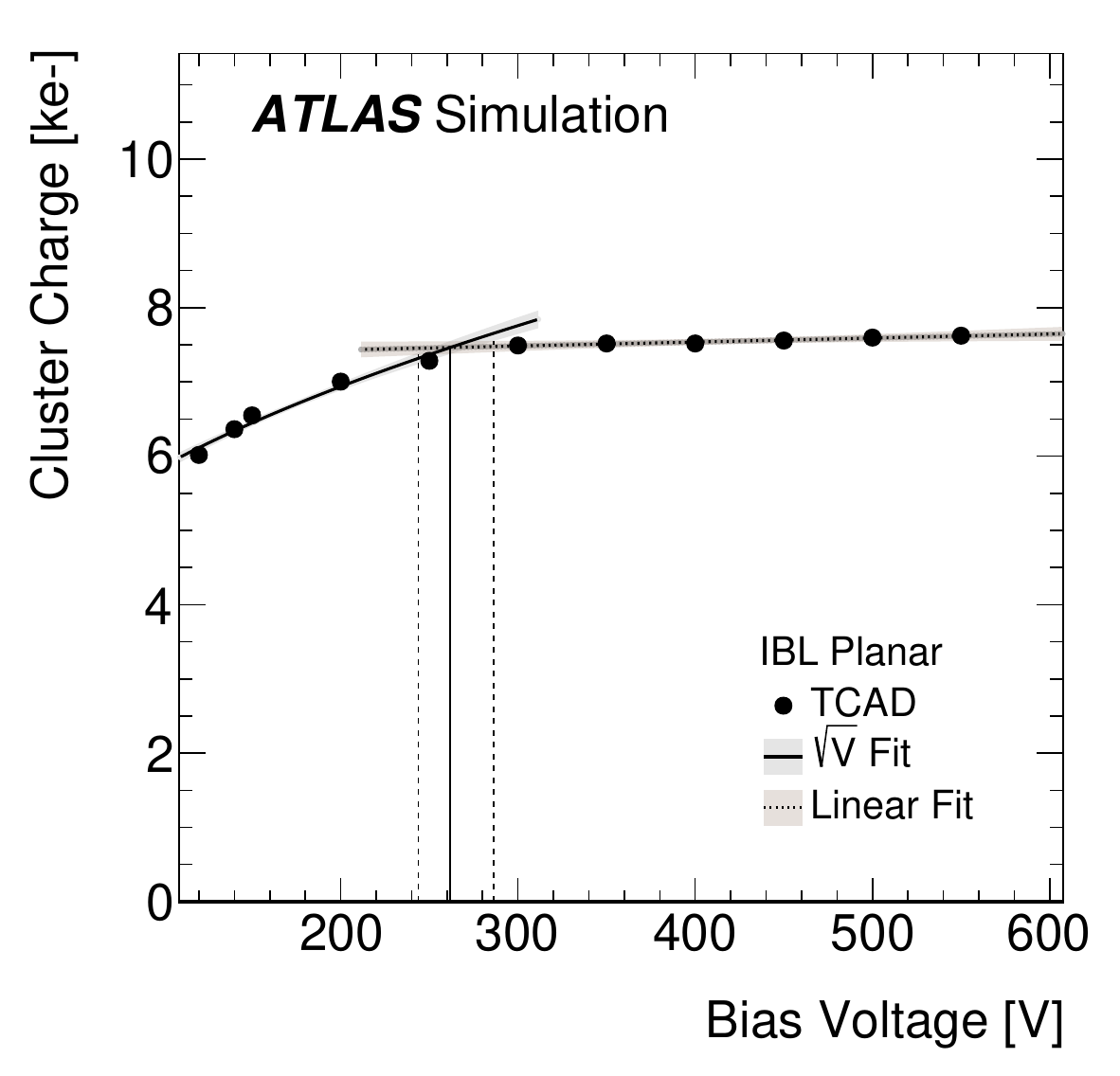} \\
\end{tabular}
\end{center}
\caption{Most probable values of the charge collected in IBL planar sensor clusters associated to reconstructed particle tracks in $pp$ collisions as a function of the applied bias voltage. The left panel shows results for data recorded at the end of 2022 after an integrated LHC delivered luminosity of 194~fb$^{-1}$ (points with error bars) and for radiation damage simulation obtained for the corresponding average fluence of $1.0 \times 10^{15}$ ~\SI{}{n_{eq}/cm^2} (grey band). The right panel shows the result for the TCAD model. \label{fig:IBLChargeVsHVData2022MCRad}}
\end{figure}

In order to investigate the charge collection properties in these conditions, a comparison of the evolution of the most probable value of the cluster charge as a function of the bias voltage applied to the sensors for this TCAD simulation, data and the full radiation damage simulation was performed at a fluence of $1.0 \times 10^{15}$ ~\SI{}{n_{eq}/cm^2}, corresponding to the average value on IBL at the end of 2022. The IBL planar pixel sensor was simulated in TCAD by implementing the charge deposition using the \texttt{singleeventupset} package of Silvaco and the collection was studied at different bias voltages. The charge values from the TCAD model are normalised to the collected charge in data at the depletion voltage. Results of this comparison are shown in Figure~\ref{fig:IBLChargeVsHVData2022MCRad}. The width of the band representing the radiation damage simulation predictions gives the systematic uncertainty in the collected charge discussed in Section~\ref{sec:Systs}, where the uncertainties are correlated bin-by-bin since the systematic contributions mostly affect the overall charge normalisation.

\begin{figure}[ht!]
\begin{center}
\includegraphics[width=.49\textwidth]{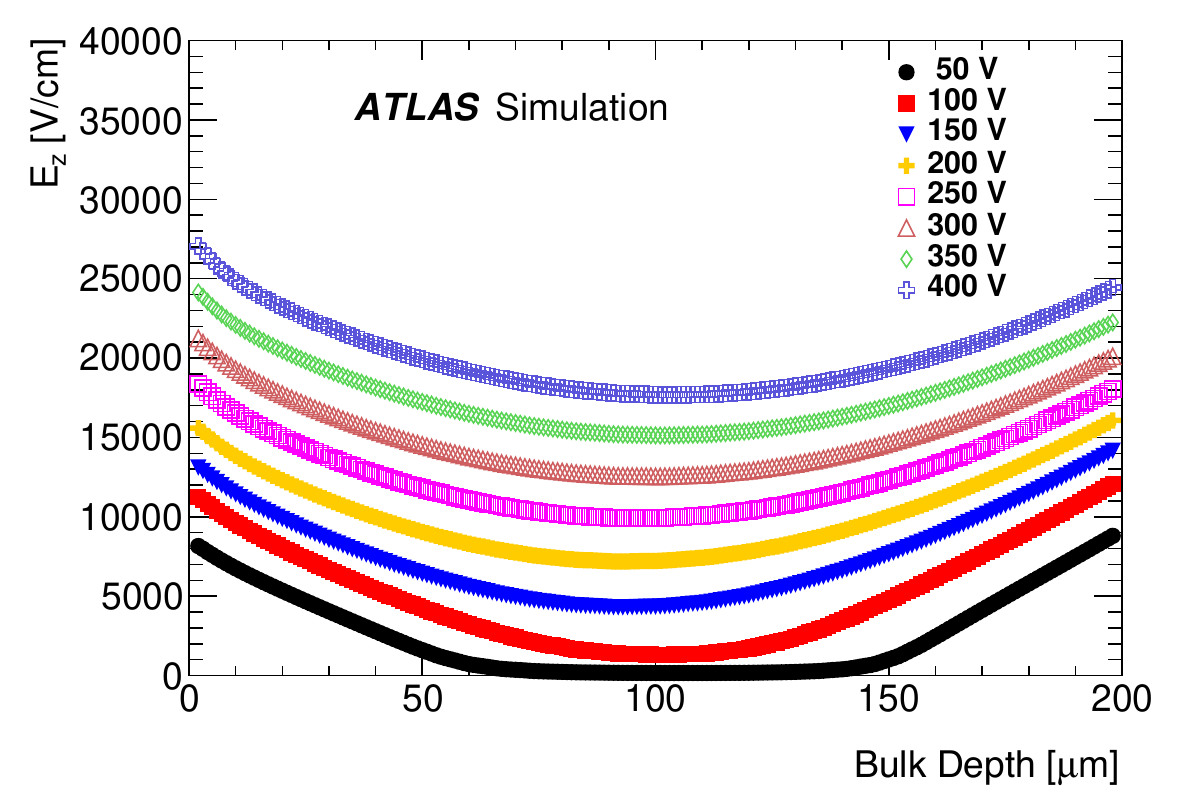} %
\end{center}
\caption{\label{fig:TCAD:E_Field}Simulated TCAD electric field  as a function of bulk depth at different bias voltages for an IBL planar pixel module irradiated at a fluence of $\Phi = 1.0 \times10^{15}$~\SI{}{n_{eq}/cm^2}.
}
\end{figure}

The collected charge in the TCAD simulation follows the same trend observed in data. In these TCAD simulations the mechanisms for charge deposition, charge trapping and signal induction are different from those used in the ATLAS radiation damage simulation. The depletion voltage is obtained from the point of transition between the $\propto \sqrt{V}$ and $\propto V$ regimes of charge variation with the bias voltage, as discussed in Section~\ref{sec:depl}. It corresponds to \SI{}{V^{data}_{depl}} = 263~$^{+27}_{-25}$~V for data, \SI{}{V^{MCRad}_{depl}} = 283~$^{+23}_{-20}$~V for the radiation damage simulation and \SI{}{V^{TCAD}_{depl}} = 262~$\pm{24}$~V for the TCAD model. The results show good agreement, despite the differences in the charge trend with voltage in the case of the TCAD model, due to the different underlying assumptions in the simulation.
This indicates that the electric field is well simulated in the TCAD simulation and that is the dominant factor in determining the observed trend of collected charge with bias voltage through the mobility. The electric field profiles at different voltages were investigated. Figure~\ref{fig:TCAD:E_Field} shows the simulated electric field at different bias voltages.%
The electric field profile shape does not change significantly when the bias voltage changes from below to above the depletion voltage, i.e.\ from 200 to 250~V under the conditions used in this simulation, but its magnitude is uniformly increased. The evolution of the TCAD simulated values of the differential capacitance as a function of the bias voltage indicates a much lower depletion voltage, around 90~V. This indicates that the role of the electric field on the collected charge is not direct but occurs through a more complex mechanism.

\begin{figure}[ht!]
\begin{center}
\begin{tabular}{cc}
\includegraphics[width=.45\textwidth]{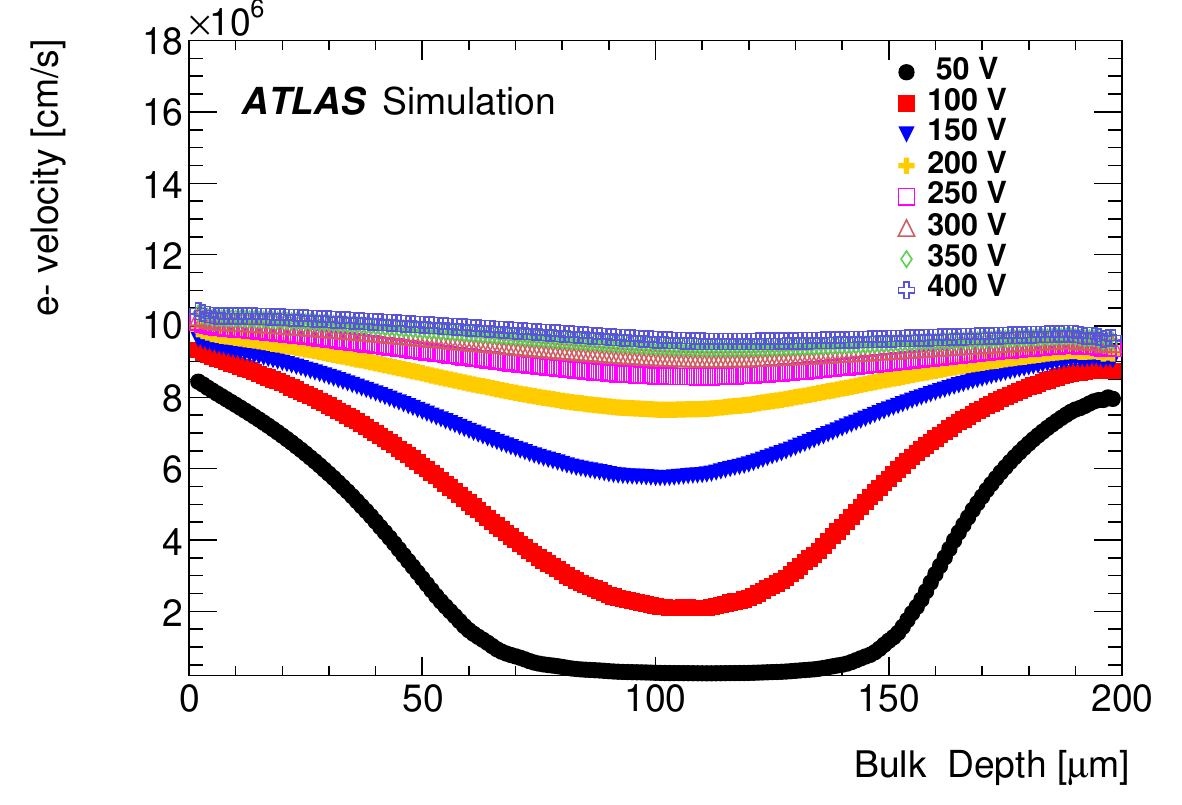} &
\includegraphics[width=.45\textwidth]{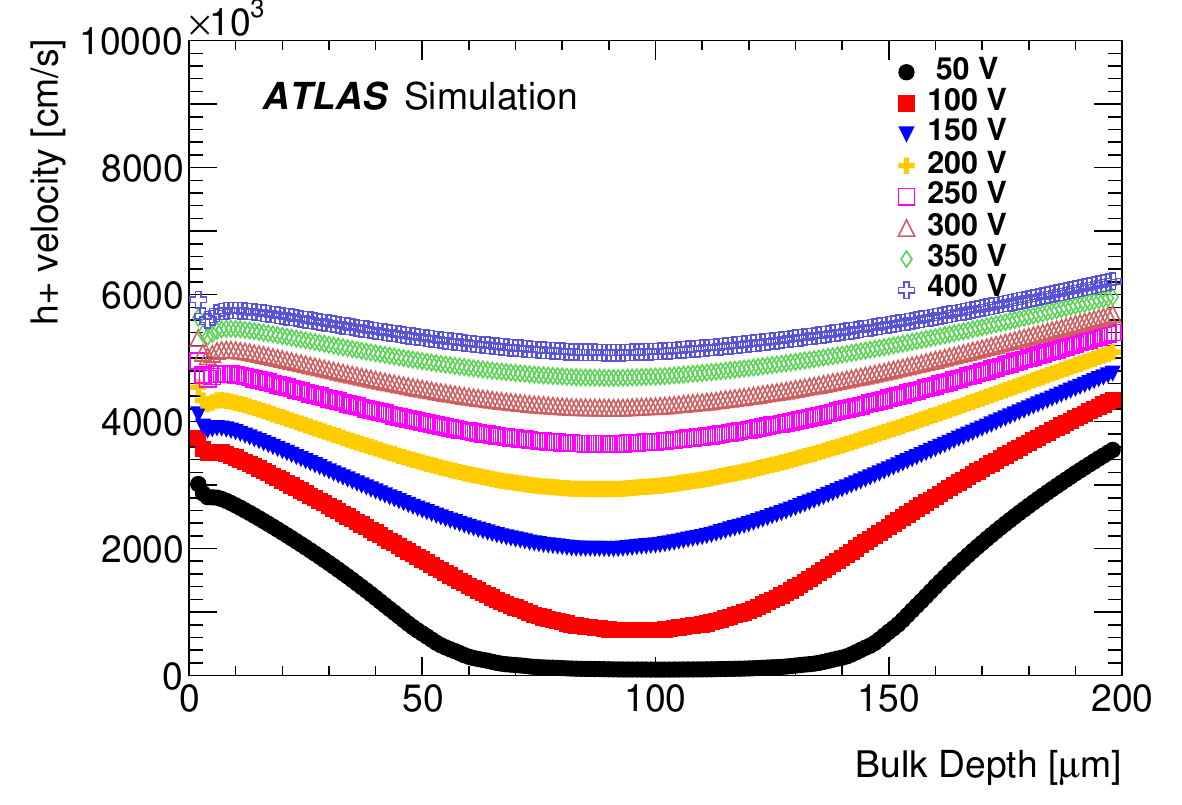} \\
\end{tabular}
\end{center}
\caption{Simulated TCAD carrier velocities for electrons (left panel) and holes (right panel) as a function of the bulk depth of charge generation at different bias voltages for an IBL planar pixel sensor irradiated at a fluence of $\Phi = 1.0 \times10^{15}$~\SI{}{n_{eq}/cm^2}. \label{fig:TCAD:velocitytimes}}
\end{figure}

This mechanism relates the change in bias voltage to the change of the electron drift velocity and of the charge carrier collection time. The length of charge carrier drift is limited by the trapping effects from radiation induced defects in the Si bulk. At $\Phi = 1.0 \times10^{15}$~\SI{}{n_{eq}/cm^2}, the trapping times for electrons and holes are of the order of 2 and 1.5~ns respectively~\cite{ATLAS:2019epb}, which are of the same order of magnitude as the collection time for a charge carrier crossing the entire sensor thickness at saturated drift velocity, as shown in Figure~\ref{fig:TCAD:velocitytimes}, where the dependence of the carrier velocity on the bulk depth was extracted from the TCAD simulation.

\begin{figure}[ht!]
\begin{center}
\includegraphics[width=.575\textwidth]{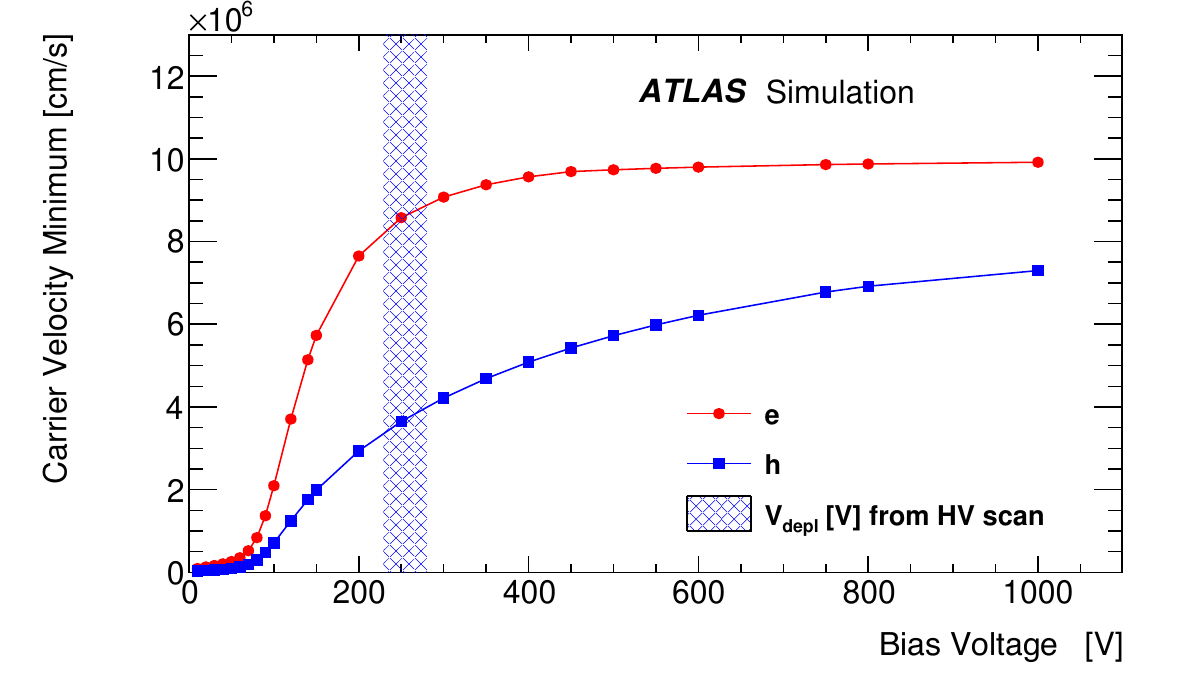}
\end{center}
\caption{\label{fig:TCAD:minVelocities}Minimum of simulated TCAD carrier velocities  as a function of  bias voltages for an IBL planar pixel module irradiated at a fluence of $\Phi = 1.0 \times10^{15}$~\SI{}{n_{eq}/cm^2}. The vertical band represents the depletion voltage measured in data.}
\end{figure}

\begin{figure}[ht!]
\begin{center}
\includegraphics[width=.495\textwidth]{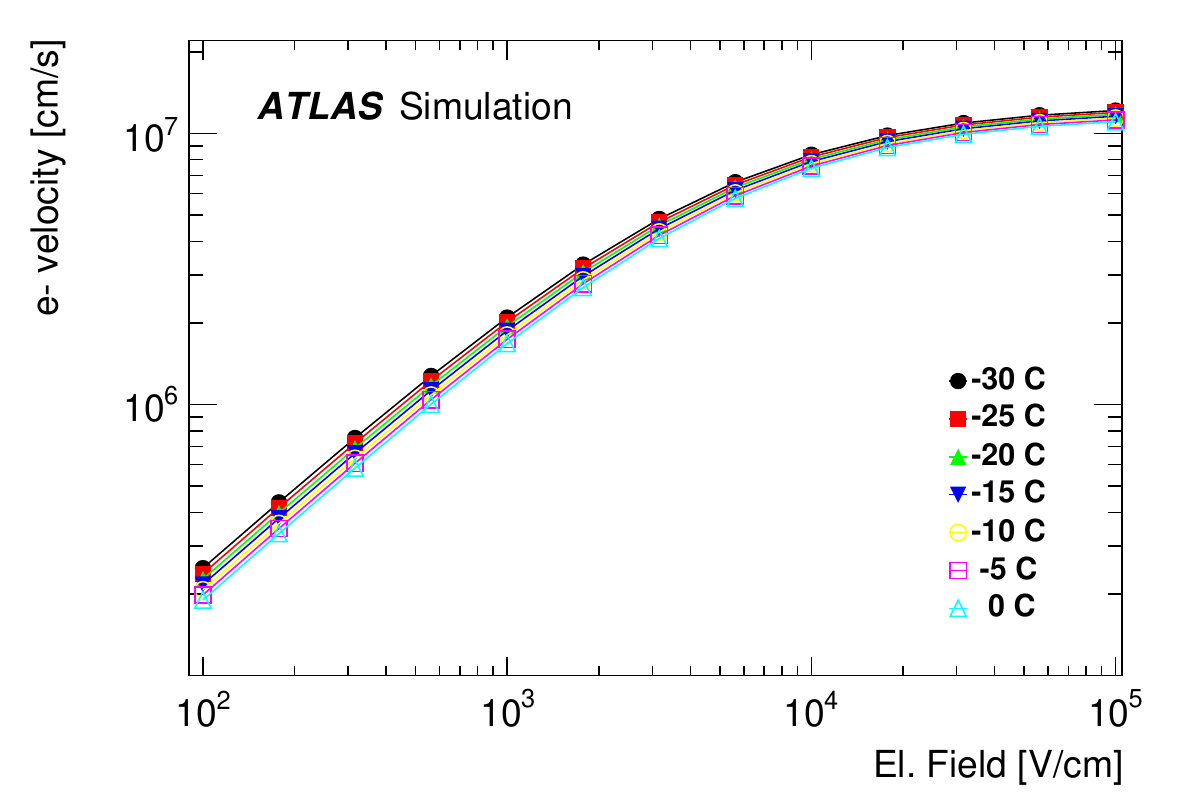}
\includegraphics[width=.495\textwidth]{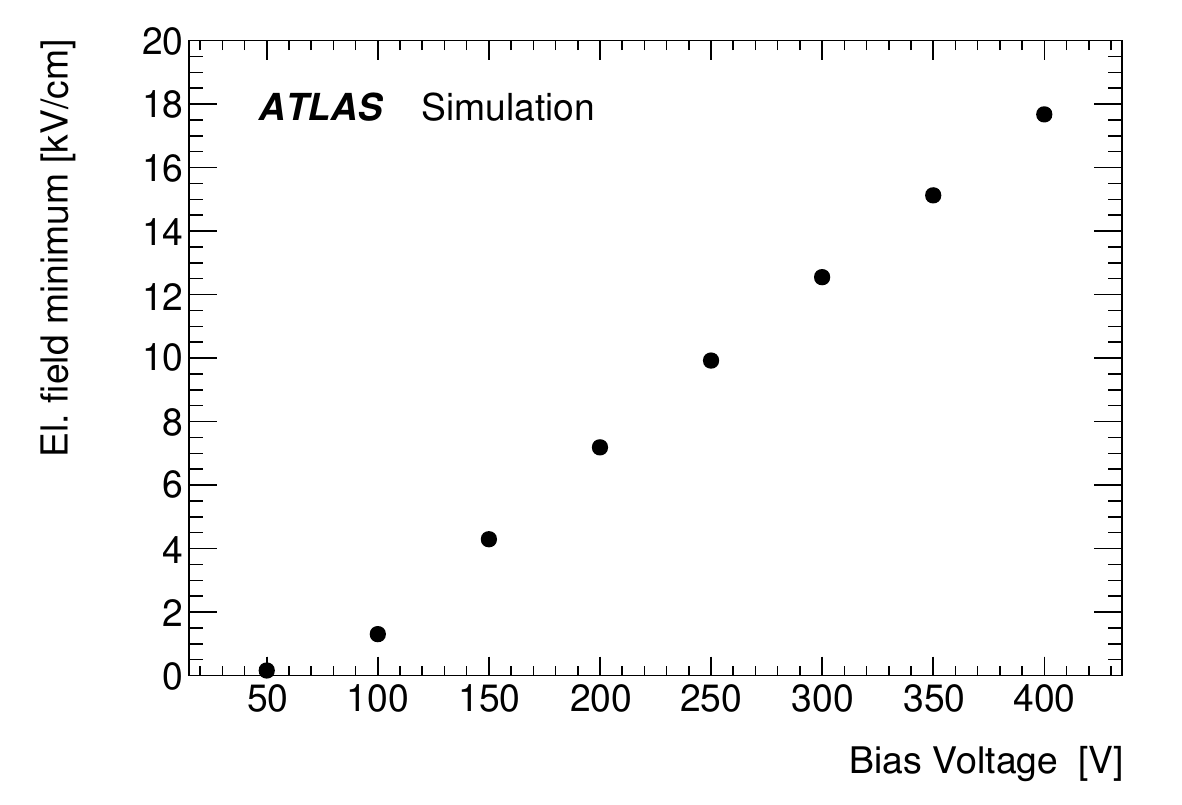}
\end{center}
\caption{\label{fig:TCAD:minElectricFieldandMobility}(Left) Electron velocity as a function of the electric field at different temperatures as predicted by the Canali model~\cite{CANALI19751122}. (Right) Minimum of simulated TCAD electric field  as a function of  bias voltages for an IBL planar pixel module irradiated at a fluence of $\Phi = 1.0 \times10^{15}$~\SI{}{n_{eq}/cm^2}.}
\end{figure}

This dependence shows that an increase of the bias voltage beyond 250~V at the reference particle fluence of $1.0 \times10^{15}$~\SI{}{n_{eq}/cm^2} has only a limited effect on the electrons drift velocities. The same is not true for the holes, but their contribution to the amplitude of the collected signal is small compared to that of electrons~\cite{rossi2006pixel}.  Using the radiation damage digitiser, the relative change in collected charge neglecting the hole contribution is estimated to be $\sim$2\%.  The carrier velocities have a local minimum in the region close to the sensor mid-plane. This minimum becomes less pronounced as the bias voltage is increased and, beyond 250~V, this feature disappears. The charge collection efficiency is lower in the regions where the carrier velocities are smaller and these observations make it interesting to investigate the evolution of the minima of the carrier velocities as a function of the bias voltage as a way to explain the corresponding evolution of the collected charge (see Figure~\ref{fig:TCAD:minVelocities}).
The increase in electron velocity has a drastic change in trend at bias voltage values corresponding to the observed change of trend in the collected charge. The origin of the observed change lies in the dependence of the mobility on the electric field. MC simulations use the so-called Canali model~\cite{CANALI19751122} and obtain good agreement with data~\cite{ATL-INDET-PUB-2018-001}. The electron velocities predicted by the Canali model are presented in the left panel of Figure~\ref{fig:TCAD:minElectricFieldandMobility}, as a function of the electric field amplitude at different temperatures.

\begin{figure}[ht!]
\begin{center}
\begin{tabular}{cc}
\includegraphics[width=0.45\textwidth]{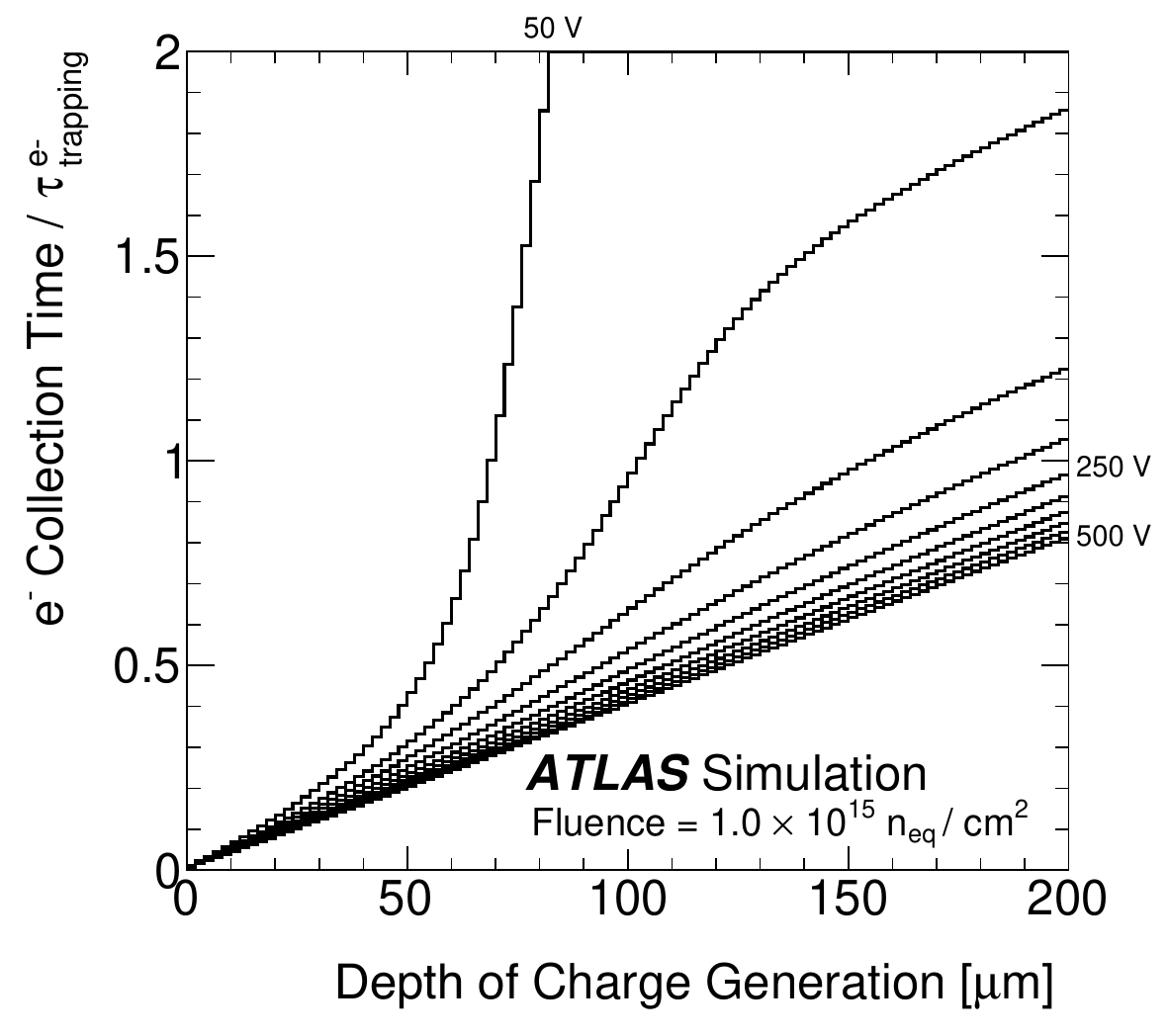}
&
\includegraphics[width=0.45\textwidth]{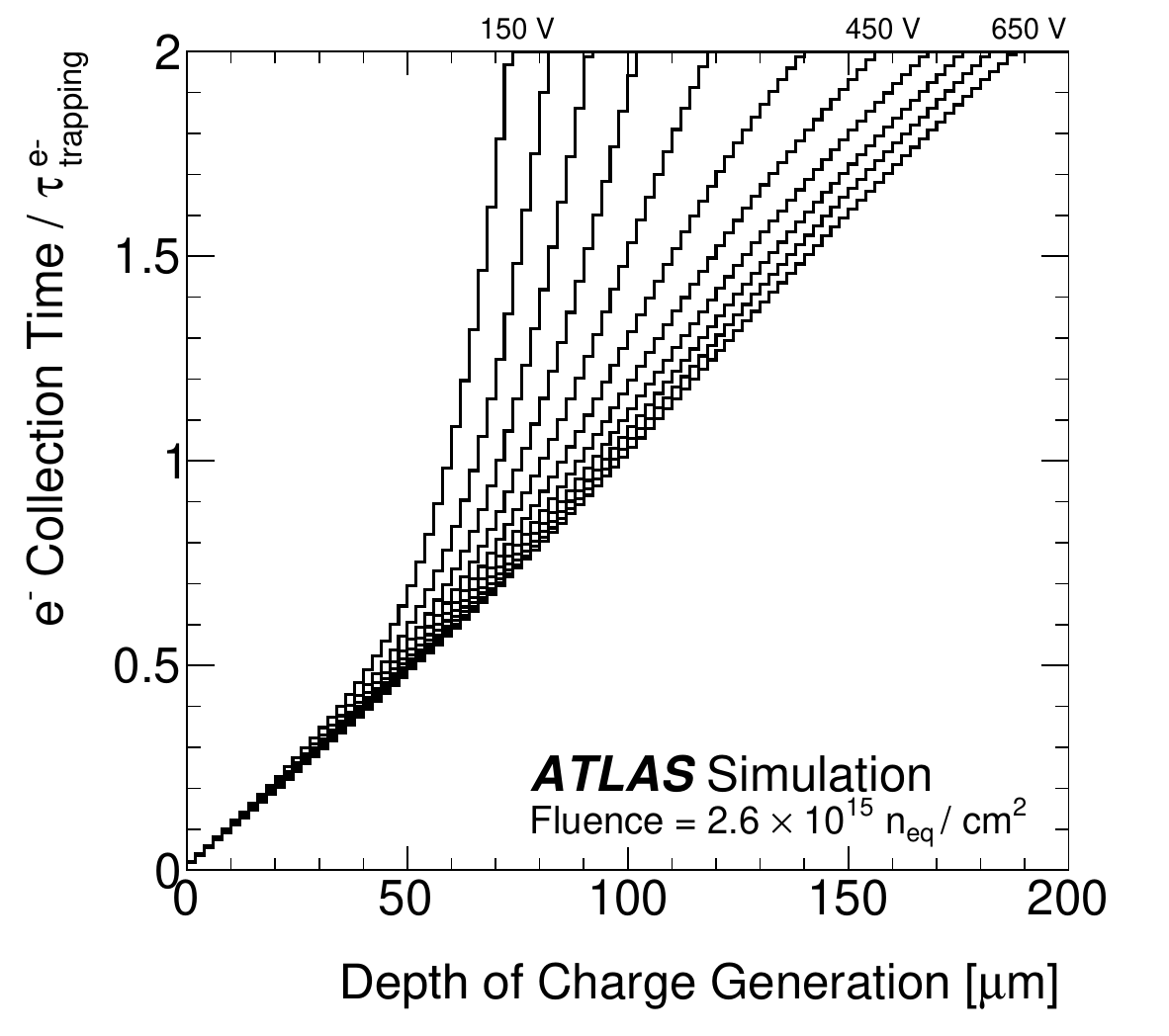} \\
\end{tabular}
\caption{\label{fig:timeToReachElectrode}
Simulated time for electrons to drift to the collecting electrode, normalised to the trapping time, as a function of the depth of the carrier creation in the Si bulk for different bias voltages, shown by the curves in steps of 50~V. Results are obtained for an IBL planar pixel sensor irradiated at a fluence of $\Phi = 1.0 \times10^{15}$~\SI{}{n_{eq}/cm^2} (left) and $2.6\times10^{15}$~\SI{}{n_{eq}/cm^2} (right). The side labels indicate values of the simulated bias voltage.}
\end{center}
\end{figure}

The increase in electron velocity has a drastic change around $10^4$~V/cm independent of the temperature. This value corresponds to the minimum of the electric field  at 250~V from the TCAD simulation.  The evolution of the minimum of the electric field with the bias voltage shown in the right panel of Figure~\ref{fig:TCAD:minElectricFieldandMobility} confirms that at around 250~V the electric field is larger than $10^4$~V/cm everywhere in the sensor bulk. This confirms that the change in the trend of increase of the collected charge with the bias voltage is due to the saturation of electron drift velocities as predicted by the mobility model. Another way to understand this effect is to study the time elapsing for charge carriers to reach the collecting electrode. This is shown as a function of the depth of the position of charge generation in Figure~\ref{fig:timeToReachElectrode}, normalised to the corresponding trapping time for different values of the applied bias voltage. The change in the functional shape of these curves reflects in the trends of increase of the collected charge with bias voltage.

For a fluence of $10^{15}$~\SI{}{n_{eq}/cm^2}, the normalised collection time deviates from a simple linear dependence with the distance from the collecting electrode when the carriers are deposited in the deeper part of the sensor, more than 100~\SI{}{\micro\meter} from the collection electrodes for electrons, at bias voltages below $\simeq$250~V. Around this value, electrons generated through the full bulk thickness have collection times shorter than the trapping time, as shown in the left panel of Figure~\ref{fig:timeToReachElectrode}. The convolution of these effects induces a departure from linearity of the collected charge with bias voltage below the depletion voltage.
For higher fluences, such as the $2.5 \times 10^{15}$~\SI{}{n_{eq}/cm^2} attained at the end of 2025, electrons produced in the deeper part of the bulk have collection times larger than the trapping times for all the bias voltage values used in the scans (up to 650~V), as shown in the right panel of Figure~\ref{fig:timeToReachElectrode}. In this regime, there is no change in the functional shape of the collected charge vs.\ bias voltage and the constant slope of the charge collection increase with voltage depends primarily on the trapping constant.

%
%
%
%
%
%
%
%
%


\section{Conclusions}
\label{sec:conclusions}

%
%

Ten years of operation at the LHC have brought a significant particle fluence on the planar pixel sensors equipping the innermost layer of the ATLAS Pixel detector. This study has discussed the evolution of the charge collection with fluence and bias voltage based on data measurements, TCAD simulations and predictions from the radiation damage digitiser in the ATLAS simulation code. Results on charge collection efficiency as a function of integrated delivered luminosity, of depth of charge generation in the Si bulk and of bias voltage show good agreement between data and simulation predictions.

The depletion voltage, defined as the point of change of the functional dependence of the collected charge increase with the applied voltage, has evolved linearly with integrated luminosity and fluence, from 110~V, after 42~fb$^{-1}$ at the end of the second year of operation in Run~2, to 500~V after 410~fb$^{-1}$ in Fall 2025. These functional dependences of the collected charge increase are understood in terms of the electron collection time as a function of the depth of charge generation and their trapping time, due to bulk damage, based on TCAD simulation studies.

At fluences of  $\sim 2.5 \times 10^{15}$~\SI{}{n_{eq}/cm^2}, electrons produced in the deeper part of the bulk have collection times larger than the trapping times for all the bias voltage values, up to 650~V, used in the voltage scans. In these conditions, the observed functional shape of the collected charge vs.\ bias voltage from these measurements is reduced to a constant slope  and the operational bias voltage of 650~V is chosen for 2026 operation, based on the resulting hits-on-track efficiency.


%
%
%
%
%
\section*{Acknowledgements}
%

%

%
%

%
%

We thank CERN for the very successful operation of the LHC and its injectors, as well as the support staff at
CERN and at our institutions worldwide without whom ATLAS could not be operated efficiently.

The crucial computing support from all WLCG partners is acknowledged gratefully, in particular from CERN, the ATLAS Tier-1 facilities at TRIUMF/SFU (Canada), NDGF (Denmark, Norway, Sweden), CC-IN2P3 (France), KIT/GridKA (Germany), INFN-CNAF (Italy), NL-T1 (Netherlands), PIC (Spain), RAL (UK) and BNL (USA), the Tier-2 facilities worldwide and large non-WLCG resource providers. Major contributors of computing resources are listed in Ref.~\cite{ATL-SOFT-PUB-2026-001}.

We gratefully acknowledge the support of ANPCyT, Argentina; YerPhI, Armenia; ARC, Australia; BMWFW and FWF, Austria; ANAS, Azerbaijan; CNPq and FAPESP, Brazil; NSERC, NRC and CFI, Canada; CERN; ANID, Chile; CAS, MOST and NSFC, China; Minciencias, Colombia; MEYS CR, Czech Republic; DNRF and DNSRC, Denmark; IN2P3-CNRS and CEA-DRF/IRFU, France; SRNSFG, Georgia; BMFTR, HGF and MPG, Germany; GSRI, Greece; RGC and Hong Kong SAR, China; ICHEP and Academy of Sciences and Humanities, Israel; INFN, Italy; MEXT and JSPS, Japan; CNRST, Morocco; NWO, Netherlands; RCN, Norway; MNiSW, Poland; FCT, Portugal; MNE/IFA, Romania; MSTDI, Serbia; MSSR, Slovakia; ARIS and MVZI, Slovenia; DSI/NRF, South Africa; MICIU/AEI, Spain; SRC and Wallenberg Foundation, Sweden; SERI, SNSF and Cantons of Bern and Geneva, Switzerland; NSTC, Taipei; TENMAK, T\"urkiye; STFC/UKRI, United Kingdom; DOE and NSF, United States of America.

Individual groups and members have received support from BCKDF, CANARIE, CRC and DRAC, Canada; CERN-CZ, FORTE and PRIMUS, Czech Republic; COST, ERC, ERDF, Horizon 2020 and Marie Sk{\l}odowska-Curie Actions, European Union; Investissements d'Avenir Labex, Investissements d'Avenir Idex and ANR, France; DFG and AvH Foundation, Germany; Herakleitos, Thales and Aristeia programmes co-financed by EU-ESF and the Greek NSRF, Greece; BSF-NSF and MINERVA, Israel; NCN and NAWA, Poland; La Caixa Banking Foundation, CERCA and AGAUR programs from Generalitat de Catalunya and PROMETEO and GenT Programmes Generalitat Valenciana, Spain; G\"{o}ran Gustafssons Stiftelse, Sweden; The Royal Society and Leverhulme Trust, United Kingdom; Eric and Wendy Schmidt Fund for Strategic Innovation, United States of America.

In addition, individual members wish to acknowledge support from Chile: Agencia Nacional de Investigaci\'on y Desarrollo (ANID FONDECYT reg. 1230987, FONDECYT 1230812, FONDECYT 1240864, Fondecyt 3240661, Fondecyt Regular 1240721); China: Chinese Ministry of Science and Technology (MOST-2023YFA1605700, MOST-2023YFA1609300), National Natural Science Foundation of China (NSFC 12275265, NSFC-W2543005); Czech Republic: Czech Science Foundation (GACR - 24-11373S), Ministry of Education Youth and Sports (ERC-CZ-LL2327, FORTE CZ.02.01.01/00/22\_008/0004632), PRIMUS Research Programme (PRIMUS/21/SCI/017); EU: H2020 European Research Council (ERC - 101002463); European Union: European Research Council (BARD No. 101116429, ERC - 948254, ERC 101089007), European Regional Development Fund (HE COFUND GA No.101081355, ERDF), Marie Sklodowska-Curie Actions (GAP-101168829); France: Agence Nationale de la Recherche (ANR-21-CE31-0013, ANR-22-EDIR-0002, ANR-24-CE31-0504-01); Germany: Deutsche Forschungsgemeinschaft (DFG - 469666862); China: Research Grants Council (GRF); Italy: Istituto Nazionale di Fisica Nucleare (LHC-MIUR - 28003/2025), Ministero dell'Università e della Ricerca (NextGenEU 153D23001490006 M4C2.1.1, NextGenEU I53D23000820006 M4C2.1.1, SOE2024\_0000023); Japan: Japan Society for the Promotion of Science (JSPS KAKENHI  JP25H0063, JSPS KAKENHI JP22H01227, JSPS KAKENHI JP22H04944, JSPS KAKENHI JP22KK0227, JSPS KAKENHI JP24K23939, JSPS KAKENHI JP24KK0251, JSPS KAKENHI JP25H00650, JSPS KAKENHI JP25H01291, JSPS KAKENHI JP25K01011, JSPS KAKENHI JP25K01023); Poland: Polish National Science Centre (NCN 2021/42/E/ST2/00350, NCN OPUS 2023/51/B/ST2/02507, NCN OPUS nr 2022/47/B/ST2/03059, NCN UMO-2019/34/E/ST2/00393, UMO-2022/47/O/ST2/00148, UMO-2023/49/B/ST2/04085, UMO-2023/51/B/ST2/00920, UMO-2024/53/N/ST2/00869); Spain: Agència de Gestió d'Ajuts Universitaris i de Recerca. (AGAUR - 2023 BP 00141), Ministry of Science and Innovation (RYC2019-028510-I, RYC2020-030254-I, RYC2021-031273-I, RYC2022-038164-I), Ministerio de Ciencia, Innovación y Universidades/Agencia Estatal de Investigaci\'on (EU NextGenerationEU (PRTR-C17.I1), PID2022-142604OB-C22); Sweden: Carl Trygger Foundation (Carl Trygger Foundation CTS 22:2312), Swedish Research Council (Swedish Research Council 2023-04654, VR 2021-03651, VR 2022-03845, VR 2022-04683, VR 2023-03403, VR 2024-05451, VR 2025-05940), Knut and Alice Wallenberg Foundation (KAW 2023.0366); Switzerland: Swiss National Science Foundation (SNSF - PCEFP2\_194658); United Kingdom: The Binks Trust, Royal Society (NIF-R1-231091); United States of America: U.S. Department of Energy (ECA DE-AC02-76SF00515), John Templeton Foundation (John Templeton Foundation 63206), Neubauer Family Foundation.

%
%


%
%
%
%
%
%

%
%
\printbibliography
\clearpage
 
\begin{flushleft}
\hypersetup{urlcolor=black}
{\Large The ATLAS Collaboration}

\bigskip

\AtlasOrcid[0000-0002-6665-4934]{G.~Aad}$^\textrm{\scriptsize 102}$,
\AtlasOrcid[0000-0001-7616-1554]{E.~Aakvaag}$^\textrm{\scriptsize 17}$,
\AtlasOrcid[0000-0002-5888-2734]{B.~Abbott}$^\textrm{\scriptsize 121}$,
\AtlasOrcid[0000-0002-0287-5869]{S.~Abdelhameed}$^\textrm{\scriptsize 83b}$,
\AtlasOrcid[0000-0002-1002-1652]{K.~Abeling}$^\textrm{\scriptsize 54}$,
\AtlasOrcid[0000-0001-5763-2760]{N.J.~Abicht}$^\textrm{\scriptsize 48}$,
\AtlasOrcid[0000-0002-8496-9294]{S.H.~Abidi}$^\textrm{\scriptsize 30}$,
\AtlasOrcid[0009-0003-6578-220X]{M.~Aboelela}$^\textrm{\scriptsize 44}$,
\AtlasOrcid[0000-0002-9987-2292]{A.~Aboulhorma}$^\textrm{\scriptsize 36e}$,
\AtlasOrcid[0000-0001-5329-6640]{H.~Abramowicz}$^\textrm{\scriptsize 154}$
% ...
\bigskip
\\

% ...
$^{17}$Department for Physics and Technology, University of Bergen, Bergen; Norway.\\
% ...
$^{30}$Physics Department, Brookhaven National Laboratory, Upton NY; United States of America.\\
% ...
$^{36}$$^{(a)}$Facult\'e des Sciences Ain Chock, Universit\'e Hassan II de Casablanca;$^{(b)}$Facult\'{e} des Sciences, Universit\'{e} Ibn-Tofail, K\'{e}nitra;$^{(c)}$Facult\'e des Sciences Semlalia, Universit\'e Cadi Ayyad, LPHEA-Marrakech;$^{(d)}$LPMR, Facult\'e des Sciences, Universit\'e Mohamed Premier, Oujda;$^{(e)}$Facult\'e des sciences, Universit\'e Mohammed V, Rabat;$^{(f)}$Institute of Applied Physics, Mohammed VI Polytechnic University, Ben Guerir; Morocco.\\
% ...
$^{44}$Physics Department, Southern Methodist University, Dallas TX; United States of America.\\
% ...
$^{48}$Fakult\"{a}t Physik, Technische Universit{\"a}t Dortmund, Dortmund; Germany.\\
% ...
$^{54}$II. Physikalisches Institut, Georg-August-Universit\"{a}t G\"ottingen, G\"ottingen; Germany.\\
% ...
$^{83}$$^{(a)}$Khalifa University of Science and Technology, Abu Dhabi;$^{(b)}$New York University Abu Dhabi, Abu Dhabi;$^{(c)}$United Arab Emirates University, Al Ain;$^{(d)}$University of Sharjah, Sharjah; United Arab Emirates.\\
% ...
$^{102}$CPPM, Aix-Marseille Universit\'e, CNRS/IN2P3, Marseille; France.\\
% ...
$^{121}$Homer L. Dodge Department of Physics and Astronomy, University of Oklahoma, Norman OK; United States of America.\\
% ...
$^{154}$Raymond and Beverly Sackler School of Physics and Astronomy, Tel Aviv University, Tel Aviv; Israel.\\
% ...
$^{a}$ Also at Affiliated with an institute formerly covered by a cooperation agreement with CERN.\\
$^{b}$ Also at An-Najah National University, Nablus; Palestine.\\
$^{c}$ Also at Borough of Manhattan Community College, City University of New York, New York NY; United States of America.\\
$^{d}$ Also at Center for Interdisciplinary Research and Innovation (CIRI-AUTH), Thessaloniki; Greece.\\
$^{e}$ Also at Centre of Physics of the Universities of Minho and Porto (CF-UM-UP); Portugal.\\
$^{f}$ Also at CERN, Geneva; Switzerland.\\
% ...

\end{flushleft}


%

%
%
%
%

%
%
%
%

\end{document}